\documentclass[aps,pre,twocolumn,superscriptaddress,noshowpacs]{revtex4}
\usepackage{amsmath}
\usepackage{amssymb}
\usepackage{tikz}
\usepackage{graphicx}% Include figure files
\usepackage{dcolumn}% Align table columns on decimal point
\usepackage{bm}% bold math
\usepackage{epsfig}
\usepackage{psfrag}
\def\II{\hbox{$1\hskip -1.2pt\vrule depth 0pt height 1.6ex width 0.7pt\vrule depth 0pt height 0.3pt width 0.12em$}}
\newcommand{\reffig}[1]{\mbox{Fig.~\ref{#1}}}
\newcommand{\refeq}[1]{\mbox{Eq.~(\ref{#1})}}
\newcommand{\refsec}[1]{\mbox{Sec.~\ref{#1}}}
\newcommand{\reftab}[1]{\mbox{Tab.~\ref{#1}}}
\newcommand{\be}{\begin{equation}}
\newcommand{\ee}{\end{equation}}
\newcommand{\bal}{\begin{align}}
\newcommand{\eal}{\end{align}}
\newcommand{\ba}{\begin{eqnarray}}
\newcommand{\ea}{\end{eqnarray}}

\newcommand{\T}{${\mathcal T}\,$}
\newcommand{\Ti}{${\mathcal T}$}
\def\II{\hbox{$1\hskip -1.2pt\vrule depth 0pt height 1.6ex width 0.7pt\vrule depth 0pt height 0.3pt width 0.12em$}}

\begin{document}

\title{\bf Properties of the eigenmodes and quantum-chaotic scattering in a superconducting microwave Dirac billiard with threefold rotational symmetry}
\author{Weihua Zhang}
\email{zhangwh2018@gmail.com}
\address{%
Lanzhou Center for Theoretical Physics and the Gansu Provincial Key Laboratory of Theoretical Physics, Lanzhou University, Lanzhou, Gansu 730000, China
}%address
\address{Center for Theoretical Physics of Complex Systems, Institute for Basic Science (IBS), Daejeon 34126, Korea}
\author{Xiaodong Zhang}
\address{%
Lanzhou Center for Theoretical Physics and the Gansu Provincial Key Laboratory of Theoretical Physics, Lanzhou University, Lanzhou, Gansu 730000, China
}%address
\author{Jiongning Che}
\address{%
Lanzhou Center for Theoretical Physics and the Gansu Provincial Key Laboratory of Theoretical Physics, Lanzhou University, Lanzhou, Gansu 730000, China
}%address
\author{M. Miski-Oglu}
\address{%
GSI Helmholtzzentrum f\"ur Schwerionenforschung GmbH, D-64291 Darmstadt, Germany}
\author{Barbara Dietz}
\email{bdietzp@gmail.com}
\address{%
Lanzhou Center for Theoretical Physics and the Gansu Provincial Key Laboratory of Theoretical Physics, Lanzhou University, Lanzhou, Gansu 730000, China
}%address
\address{Center for Theoretical Physics of Complex Systems, Institute for Basic Science (IBS), Daejeon 34126, Korea}

\date{\today}

\bigskip

\begin{abstract}We report on experimental studies that were performed with a microwave Dirac billiard (DB), that is, a flat resonator containing metallic cylinders arranged on a triangular grid, whose shape has a threefold rotational ($C_3$) symmetry. Its band structure exhibits two Dirac points (DPs) that are separated by a nearly flat band. We present a procedure which we employed to identify eigenfrequencies and to separate the eigenstates according to their transformation properties under rotation by $\frac{2\pi}{3}$ into the three $C_3$ subspaces. This allows us to verify previous numerical results of Ref. [W. Zhang and B. Dietz, Phys. Rev. B {\bf 104}, 064310 (2021)], thus confirming that the properties of the eigenmodes coincide with those of artificial graphene around the lower DP, and are well described by a tight-binding model (TBM) for a honeycomb-kagome lattice of corresponding shape. Above all, we investigate properties of the wave-function components in terms of the fluctuation properties of the measured scattering matrix, which are numerically not accessible. They are compared to random-matrix theory predictions for quantum-chaotic scattering systems exhibiting extended or localized states in the interaction region, that is, the DB. Even in regions, where the wave functions are localized, the spectral properties coincide with those of typical quantum systems with chaotic classical counterpart. 
\end{abstract}
\bigskip
\maketitle

\section{Introduction\label{Intr}} 
\begin{figure}[!th]
\includegraphics[width=0.44\linewidth]{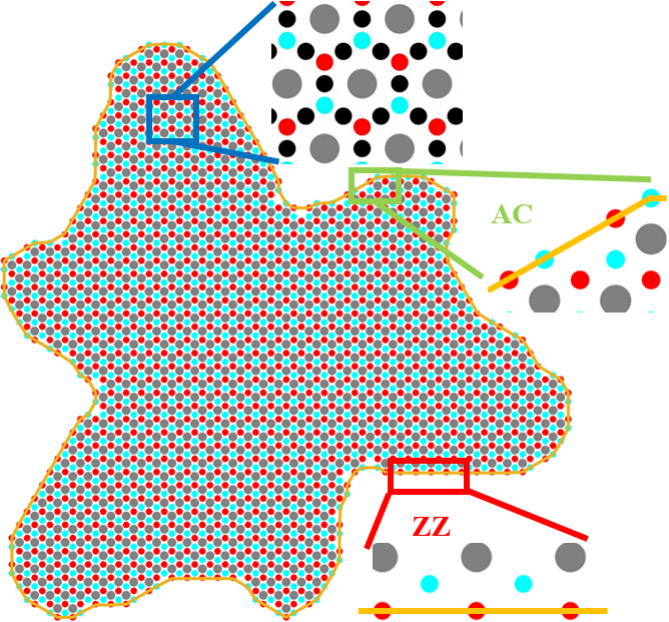}
\includegraphics[width=0.54\linewidth]{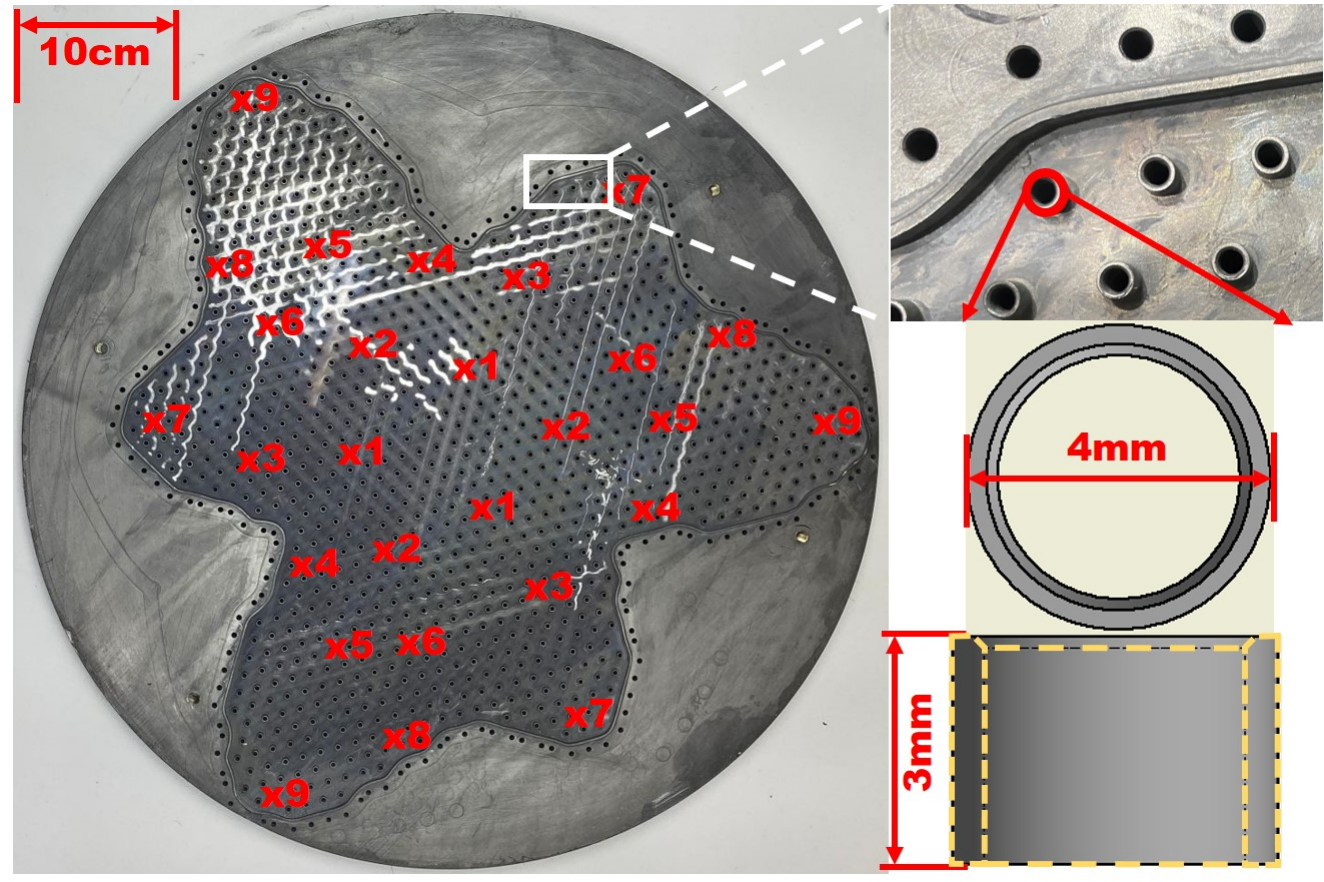}
\caption{Left panel: Schematic view of the Dirac billiard, which comprises 1033 metallic cylinders (gray disks) arranged on a triangular grid. In the upmost inset, red and turquoise dots indicate the positions of the voids. They are located on the interpenetrating triangular sublattices of the honeycomb lattice which is terminated by zigzag (ZZ) and armchair edges (AC), as indicated in the lower insets. The centers between two neighboring cylinders, marked by black dots, form a kagome structure. Right panel:  Photograph of the basin of the resonator. The metallic cylinders are milled out of a circular brass plate with radius $R=$570~mm and height 19.5~mm. The red numbers denote the nine groups of, respectively, three antennas. To achieve superconductivity, the basin and the lid, which is a circular brass plate of radius $R$ and height 6~mm with screw holes at the positions of the cylinders and along the boundary, are covered with a lead coating, whose critical temperature is $T_c$=7.2~K, and then tightly screwed together through all holes. The resonator was cooled down to 4-6~K in a cryogenic chamber constructed by ULVAC Cryogenics in Kyoto, Japan. The inset to the right shows a zoom into one of the cylinders of diameter 4~mm and height 3~mm. The upper part is designed with a cut edge shape, as indicated by the yellow dashed lines, to achieve good electrical contact with the lid~\cite{Dietz2015}.}
        \label{fig:Sketch_Diracbilliard}
\end{figure}
Superconducting microwave Dirac billiards (DBs) have been used since more than a decade to investigate fluctuation properties in the energy spectra of artificial graphene and fullerene structures~\cite{Bittner2010,Bittner2012,Dietz2013,Dietz2015,Iachello2015,Dietz2015a,Dietz2015b,Dietz2016,Dietz2019a}. The experiments presented in this work were performed with the DB shown schematically in~\reffig{fig:Sketch_Diracbilliard}, whose shape has a $C_3$ symmetry. The frequency was restricted to the range of the lowest transverse-magnetic (TM) mode, where the electric-field strength is perpendicular to the resonator plane and thus is governed by the scalar Helmholtz equation with Dirichlet boundary conditions (BCs) at the sidewalls of the cavity and cylinders. The Helmholtz equation is mathematically identical to the Schr\"odinger equation of a quantum billiard (QB) of corresponding shape, into which scatterers are inserted at the positions of the cylinders. The crucial advantage of such resonators as compared to honeycomb structures constructed from dielectric disks~\cite{Kuhl2010} is that superconducting high-precision measurements can be performed, which is indispensable for the determination of complete sequences of resonance frequencies. 

The band structure of propagating modes of the DB exhibits two Dirac points (DPs), where the first and second, respectively, the fourth and fifth band touch each other conically, and a nearly flat third band (FB) in between. It is reminiscent of that of a honeycomb-kagome billiard (HKB) whose sites form a combination of a honeycomb and a kagome sublattice~\cite{Jacqmin2014,Lan2012,Lu2017a,Zhong2017,Maimaiti2020,Zhang2021}; see the upmost inset of~\reffig{fig:Sketch_Diracbilliard}. Indeed, below the FB the electric-field intensities are maximal at the voids, that are located at the centers of three neighboring metallic cylinders (grey disks), marked with red and turquoise dots in~\reffig{fig:Sketch_Diracbilliard} and form a honeycomb structure. In the frequency range of the FB they are maximal at the centers between adjacent cylinders, marked by black dots, that are at the sites of a kagome lattice, and above the FB on all sites of the HKB~\cite{Maimaiti2020,Zhang2021}. We demonstrated that below the FB the properties of DBs are well captured by a tight-binding model (TBM) for a graphene billiard (GB)~\cite{Dietz2015,Dietz2015b,Dietz2016}, and generally by one for a HKB~\cite{Maimaiti2020,Zhang2021}. Dirac points are a characteristic of graphene, that attracted a lot of attention~\cite{Novoselov2004,Beenakker2008,Neto2009} because in the region of the conical valleys graphene features relativistic phenomena~\cite{DiVincenzo1984,Novoselov2004,Geim2007,Avouris2007,Miao2007,Ponomarenko2008,Beenakker2008,Zhang2008,Neto2009,Abergel2010,Zandbergen2010}, which triggered numerous realizations~\cite{Polini2013} of artificial graphene~\cite{Parimi2004,Joannopoulos2008,Bittner2010,Kuhl2010,Singha2011,Nadvornik2012,Gomes2012,Tarruell2012,Uehlinger2013,Rechtsmann2013,Rechtsmann2013a,Khanikaev2013,Wang2014a,Shi2015,Bellec2013,Bellec2013a,Bellec2014}. In the vicinity of the band edges (BEs) the spectral properties coincide with those of a nonrelativistic QB of corresponding shape~\cite{Dietz2015,Dietz2013,Dietz2016}. 

The classical dynamics of a billiard with the shape of the DB shown in~\reffig{fig:Sketch_Diracbilliard} is chaotic~\cite{Dembowski2000,Zhang2021}. According to the Bohigas-Giannoni-Schmit conjecture the fluctuation properties in the energy spectra of nonrelativistic quantum systems with a chaotic classical counterpart are universal~\cite{Berry1977a,Berry1979,Casati1980,Bohigas1984} and coincide with those of random matrices from the Gaussian orthogonal ensemble (GOE) for time-reversal (\Ti) invariant systems and the Gaussian unitary ensemble (GUE) if \T invariance is violated. Yet there also exist billiards with certain shapes which do not comply with this conjecture. Examples are billiards whose shape has $C_3$ symmetry~\cite{Leyvraz1996,Dembowski2000,Dembowski2003}, a unidirectional classical dynamics~\cite{Knill1998,Gutkin2007,Veble2007,Dietz2014} or nanoelectromechanical systems consisting of a circular quantum dot on a suspended nanoscopic dielectric plate~\cite{Rego2005,Gusso2006}. Their spectral properties may coincide with those of generic chaotic systems with violated time-reversal invariance even though it is preserved. The boundary of the DB has a $C_3$ symmetry. We, actually, chose the same shape as in the experiments which were performed 20 years ago with a superconducting microwave billiard in the range below the cutoff frequency $f^{cut}$ of the first transverse-electric mode, to investigate the spectral properties of the corresponding quantum billiard~\cite{Dembowski2000,Dembowski2003}. Interest in this QB arose due to theoretical predictions~\cite{Robbins1989,Leyvraz1996,Keating1997,Joyner2012} that the spectral properties of part of the spectrum coincide with those of random matrices from the GUE. The origin of these discrepancies are outlined in~\refsec{Rev}.

The objective of~\cite{Zhang2021} and the present work was the numerical and experimental study of the properties of DBs and corresponding GBs and HKBs, whose boundary has a $C_3$ symmetry, especially in the relativistic region aorund the DPs. In the region of the conical valleys, that are located on, respectively, three of the corners of the first Brillouin zone~\cite{Wallace1947}, the two sets of valley eigenstates are well described by Dirac Hamiltonians for massless spin-1/2 quasiparticles~\cite{Beenakker2008,Neto2009}. Therefore, we also investigated in Ref.~\cite{Zhang2021} properties of relativistic neutrino billiards (NBs) of corresponding shape. They were introduced in Ref.~\cite{Berry1987}, and are governed by the Weyl equation~\cite{Weyl1929} for a spin-1/2 particle. The associated Dirac Hamiltonian is not invariant under time reversal, so the spectral properties of NBs with the shape of a chaotic billiard typically coincide with those of random matrices from the GUE, if the shape has no geometric symmetries. It has been demonstrated in Refs.~\cite{Yupei2016,Yupei2022} that the spectral properties of GBs and NBs of corresponding shape do not coincide~\cite{Silvestrov2007,Ponomarenko2008,Libisch2009,Wurm2009,Huang2010,Wurm2011,Rycerz2012,Rycerz2013,Polini2013,Dietz2015,Dietz2016}. These discrepancies were attributed to intervalley scattering at the boundary of GBs~\cite{Wurm2009,Rycerz2012,Rycerz2013}. Similar observations were made for HKBs~\cite{Maimaiti2020,Zhang2021}.

In this work we present experimental results for the DB shown in~\reffig{fig:Sketch_Diracbilliard}. In~\refsec{Rev} we briefly review the properties of the billiard systems that were investigated in~\cite{Zhang2021} and the results. Then, in~\refsec{DB} we provide information on the DB and experiment. Properties of the eigenmodes~\cite{Zhang2021} are analyzed in~\refsec{Spectr} and compared to those of the corresponding GB, QB and NB. For the first time, we also analyzed fluctuation properties of the scattering ($S$) matrix describing the measurement process~\cite{Albeverio1996}. Furthermore, we investigated strength distributions~\cite{Dembowski2005,Dietz2006a} which give information on the product of wave function components at the positions of the antennas, and thus on their intensity distribution, and demonstrate that they provide a tool to detect localization, i.e., scarred wave functions, as outlined in~\refsec{Scatt}. Finally, in~\refsec{Concl} we discuss and evaluate the results.  

\section{Review of the theoretical and numerical results\label{Rev}} The domain $\Omega$ of the DB shown in~\reffig{fig:Sketch_Diracbilliard} is defined in the complex plane $w(r,\phi)=x(r,\phi)+iy(r,\phi)$ with $\phi\in [0,2\pi),\, r=[0,r_0]$ by the parametrization
\be
w(r,\phi)=r\left[1+0.2\cos(3\phi)-0.2\sin(6\phi)\right]e^{i\phi}.
\label{coordinates}
\ee
The boundary $\partial\Omega$ is given by $w(r=r_0,\phi)$. The eigenfunctions $\psi(r,\phi)$ of the QB with this shape and the electric-field strength of the corresponding microwave billiard below $f^{cut}$~\cite{Stoeckmann1990,Sridhar1991,Graef1992} are governed by the Schr\"odinger equation with Dirichlet BCs along $\partial\Omega$. The solutions can be separated into the three irreducible subspaces associated with the $C_3$ symmetry, which are defined by the transformation properties of the eigenfunctions under rotation by $\frac{2l\pi}{3},\, l=0,1,2$. The rotation operator is given by
\be
\hat R=e^{i\frac{2\pi}{3}\hat L}
\ee
with $\hat L$ denoting the angular momentum operator. Applying it to the eigenfunctions of the QB yields for the symmetry-projected ones
\be
\hat R^{\lambda}\psi^{(l)}_m(r,\phi)=\psi_m^{(l)}\left(r,\phi-\frac{2\pi}{3}\lambda\right)=e^{i\frac{2l\pi}{3}\lambda}\psi^{(l)}_m(r,\phi),
\label{RotSym}
\ee
where
\be
[\hat R,\hat H]=0.
\ee
For $l=0$ the wave functions are real and rotationally invariant, and thus invariant under the time-reversal operator $\hat T= \mathcal{\hat C}$ with $\mathcal{\hat C}$ denoting the complex conjugation operator~\cite{Haake2018}. In contrast, for $l=1,2$ they are complex and 
\be
\hat T\psi^{(1,2)}_m(r,\phi)=\psi^{(2,1)}_m(r,\phi),\label{Top}
\ee
implying that $\psi_m^{(1)}(r,\phi)$ and $\psi_m^{(2)}(r,\phi)$ are eigenfunctions with the same eigenvalue $k_m^2$. Thus, the eigenvalue spectrum can be separated into nondegenerate eigenvalues (singlets) and pairwise degenerate ones (doublets). If the corresponding classical dynamics is chaotic and if the billiard boundary has no additional symmetries, the spectral properties of the singlets show GOE behavior, while those of the two doublet partners exhibit GUE statistics~\cite{Leyvraz1996}. 

\begin{figure}[!th]
\includegraphics[width=0.6\linewidth]{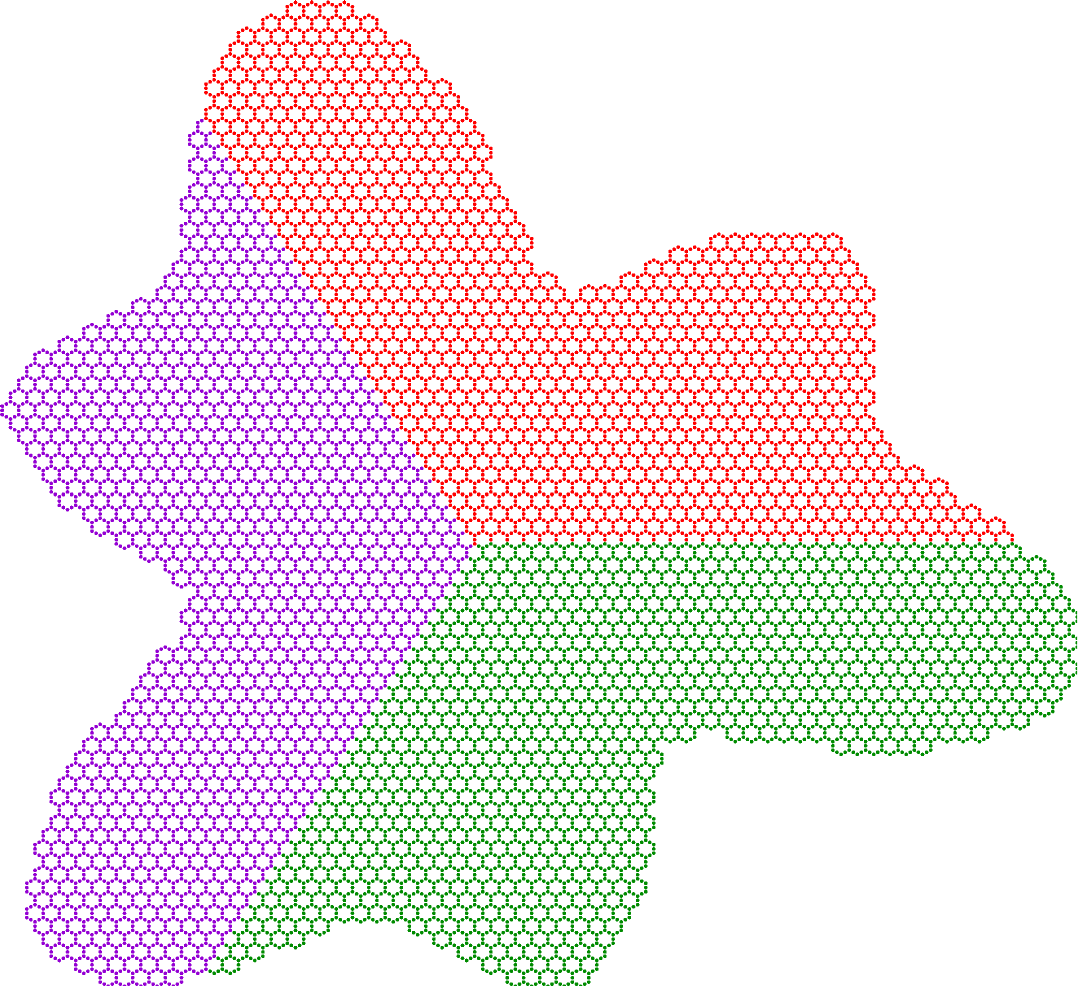}
	\caption{Illustration of the procedure used to construct the GB and HKB. They are obtained by rotating a wegde with the shape of a fundamental domain, e.g., the red one, twice around its tip yielding the purple and green ones.}
\label{Sketch}
\end{figure}

Similarly, the eigenstates of GBs and HKBs with $C_3$ symmetry can be classified according to their tranformation properties under rotation by  $\frac{2\pi}{3}$. The matrix elements of the associated TBM Hamiltonian are given by
\be
\nonumber\mathcal{\hat H}^{TBM}_{ij}=t_0\delta_{ij}+t_1\hat\delta(\vert\boldsymbol{r}_{i}-\boldsymbol{r}_{j}\vert -d_0)
+t_2\hat\delta(\vert\boldsymbol{r}_{i}-\boldsymbol{r}_{j}\vert -d_1),
\ee
where $\hat\delta(x)$ equals unity for $x=0$ and is zero otherwise, $\boldsymbol{r}_{i}$ denotes the position of site $i$, and $d_0=a_L/\sqrt{3},\, d_1=0$ for the honeycomb lattice, respectively, $d_0=a_L/(2\sqrt{3}),\,d_1=a_L/2$ for the honeycomb-kagome lattice. We constructed the GB and HKB by rotating a wedge with inner angle $\frac{2\pi}{3}$  about its tip\, as illustrated in~\reffig{Sketch}. 
The corresponding TBM Hamiltonian is $3N\times 3N$ dimensional, if each wedge comprises $N$ sites and is given by
\be
\mathcal{\hat H}_{TBM}=
\begin{pmatrix}
\hat H&\hat V&\hat V^T\\
\hat V^T&\hat H&\hat V\\
\hat V&\hat V^T&\hat H
\end{pmatrix},
\label{MHTBM}
\ee
where $\hat H$ denotes the $N$-dimensional TBM Hamiltonian of the wedge-shaped lattice structure, which is the same for each subdomain in~\reffig{Sketch}. The $N\times N$ coupling matrix $\hat V$ and its transpose $\hat V^T$ contain the hoppings between sites of two adjacent subdomains along their common boundary. The TBM Hamiltonian can be brought to block-diagonal form by applying a unitary transformation,  
\ba
\label{TBMH}
&&\hat U^\dagger\mathcal{\hat H}_{TB}\hat U=
\begin{pmatrix}
\hat H^{TB(0)} &0_N&\hat 0_N\\
\hat 0_N&\hat H^{TB(1)}&\hat 0_N\\
\hat 0_N&\hat 0_N&\hat H^{TB(2)}
\end{pmatrix},\\
&&\hat H^{TB(0)}=\hat H+\hat V+\hat V^T,\nonumber\\
&&\hat H^{TB(1)}=\hat H+e^{i\frac{2\pi}{3}}\hat V+e^{i\frac{4\pi}{3}}\hat V^T, \nonumber\\
&&\hat H^{TB(2)}=\hat H+e^{i\frac{4\pi}{3}}\hat V+e^{i\frac{2\pi}{3}}\hat V^T, \nonumber
\ea
with
\be
\hat U=\frac{1}{\sqrt{3}}
\begin{pmatrix}
\II_N&e^{i\frac{4\pi}{3}}\II_N&e^{i\frac{4\pi}{3}}\II_N\\
\II_N&\II_N&e^{i\frac{2\pi}{3}}\II_N\\
\II_N&e^{i\frac{2\pi}{3}}\II_N&\II_N
\end{pmatrix},
\ee
where $\II_N$ denotes the $N$-dimensional unit matrix. The Hamiltonians $\hat H^{TB(l)},\, l=0,1,2$ are associated with the three irreducible $C_3$ subspaces defined by the transformation properties~\refeq{RotSym} under rotation by $\frac{2\pi}{3}$.

In contrast, the spinor eigenfunctions of the corresponding NB can not be classified according to their transformation properties under rotation by $\frac{2\pi}{3}$~\cite{Zhang2021}. This is only possible for each component separately. Neutrino billiards were introduced in~\cite{Berry1987}. They are governed by the Weyl equation~\cite{Weyl1929} for a non-interacting spin-1/2 particle of mass $m_0$, which is referred to as Dirac-equation in~\cite{Berry1987} and, generally, in the context of NBs. In the two-dimensional plane $\boldsymbol{r}=(x,y)$ it is given by
\be
\boldsymbol{\hat H}_D\boldsymbol{\psi}=\left( c\boldsymbol{\hat\sigma}\cdot\boldsymbol{\hat p}+m_0c^2{\hat\sigma_z}\right)\boldsymbol{\psi}
=E\boldsymbol{\psi},\, \boldsymbol{\psi}=
\begin{pmatrix}
\psi_1 \\ \psi_2
\end{pmatrix},
\label{DE}
\ee
with $\boldsymbol{\hat p}=-i\hbar\boldsymbol{\nabla}$ the momentum of the particle. Furthermore, $\boldsymbol{\hat H}_D$ denotes the Dirac Hamiltionian, $\boldsymbol{\hat\sigma}=({\hat\sigma}_x,{\hat\sigma}_y)$, ${\hat\sigma}_{x,y,z}$ are the Pauli matrices and $E=\hbar ck_E=\hbar ck\sqrt{1+\beta^2}$ is the energy of the particle. Here $k$ is the free-space wave vector and $\beta=\frac{m_0c}{\hbar k}$ is the ratio of the rest-energy momentum and free-space momentum. In Ref.~\cite{Berry1987} only the ultrarelativistic, i.e., massless case $m_0=0$, was considered. The particle is confined to the billiard domain $\Omega$ by imposing the boundary condition that the normal component of the local current, which is given by the expectation value of the current operator $\boldsymbol{\hat u}=\boldsymbol{\nabla}_{\boldsymbol{p}}\boldsymbol{\hat H}_D=c\boldsymbol{\hat\sigma}$, $\boldsymbol{u}(\boldsymbol{r})=c\boldsymbol{\psi}^\dagger\boldsymbol{\hat\sigma}\boldsymbol{\psi}$, vanishes, yielding independently of the mass~\cite{Berry1987,Dietz2020},
\be
\psi_2(\phi)=i\mu e^{i\alpha(\phi)}\psi_1(\phi),
\label{BC1}
\ee
where $\alpha(\phi)$ is the angle of the outward-pointing normal vector $\boldsymbol{n}(\phi)$ at $w(r_0,\phi)$ with respect to the $x$ axis, and $\mu=\pm 1$ determines the rotational direction of the current at the boundary. We set it to unity in the calculations presented in Ref.~\cite{Zhang2021}. The nonrelativistic limit is reached when the energy is close to the rest energy, $E\simeq m_0c^2$~\cite{Baym2018}, that is, for sufficiently large $\beta\to\infty$. 

Like in the nonrelativistic limit~\refeq{RotSym}, the eigenstates of an NB with $C_3$ symmetry can be grouped into three subspaces defined by their transformation properties under a rotation by $\frac{2\pi}{3}$~\cite{Leyvraz1996,Keating1997,Robbins1989,Joyner2012}, yielding the symmetry-projected eigenstates
\be
\boldsymbol{\hat R}^{\lambda}\psi^{(l)}_{1,2}(\boldsymbol{r})
=e^{i\lambda\frac{2l\pi}{3}}\psi^{(l)}_{1,2}(\boldsymbol{r}), \lambda=0,1,2.
\ee
However, for a given eigenwavenumber $k_m$ the spinor components of the corresponding eigenfunctions behave differently under rotation by $\frac{2\pi}{3}$~\cite{Dietz2021,Zhang2021}. Namely, if the first component belongs to the subspace $l$,
\be
\hat R\psi_{1,m}^{(l)}(\boldsymbol{r})=e^{il\frac{2\pi}{3}}\psi_{1,m}^{(l)}(\boldsymbol{r}),\label{sympsi1}
\ee
then the Dirac equation yields for the second one
\be
\hat R\psi_{2,m}(\boldsymbol{r})=e^{i(l-1)\frac{2\pi}{3}}\psi_{2,m}(\boldsymbol{r}),\label{sympsi2}
\ee
where $l=-1$ corresponds to $l=2$. Similarly, employing~\refeq{sympsi1} in the BC~\refeq{BC1} and the $C_3$ symmetry of the boundary, that is, $e^{i\alpha\left(\phi-\lambda\frac{2\pi}{3}\right)}=e^{-i\lambda\frac{2\pi}{3}}e^{i\alpha(\phi)}$, gives~\cite{Zhang2021,Dietz2021}
\be
\label{BCm2}
\hat R\psi_{2,m}(\phi)=ie^{i\alpha\left(\phi-\frac{2\pi}{3}\right)}\psi_{1,m}^{(l)}\left(\phi-\frac{2\pi}{3}\right)
=e^{i(l-1)\frac{2\pi}{3}}\psi_{2,m}(\phi)\nonumber
\ee
implying that $\psi_{2,m}(\phi)=\psi^{(l-1)}_{2,m}(\phi)$ if $\psi_{1,m}(\phi)=\psi^{(l)}_{1,m}(\phi)$ meaning that, if the first component belongs to the subspace $l$, then the second one belongs to the subspace $(l-1)$. This intermingling of symmetry properties has its origin in the additional spin degree of freedom~\cite{Zhang2021,Dietz2021}. Nevertheless, the spinor components can be classified according to the symmetry class of, e.g., the first component and, accordingly, their eigenvalues can be assigned to symmetry-projected subspectra. In distinction to nonrelativistic QBs, the spectral properties are well described by the GUE for all subspaces, if the NB has the shape of a billiard with chaotic dynamics and no mirror symmetries.

In~\cite{Zhang2021} we computed the eigenvalues of the QBs and NBs for each symmetry class separately by employing boundary integral equations resulting from Green's theorem~\cite{Berry1987,Baecker2003,Dietz2020,Dietz2022a}. The eigenvalues of the GB and HKB were obtained by diagonalizing each block of the TBM Hamiltonian~\refeq{TBMH} separately. Furthermore, we computed with COMSOL Multiphysics the symmetry-projected resonance frequencies and electric-field distributions of the DB. For the DB, GB, HKB and QB the spectral properties of the singlets exhibit GOE statistics, those of the doublets GUE statistics~\cite{Leyvraz1996,Keating1996,Braun2011,Dembowski2000,Dembowski2003,Schaefer2003,Robbins1988,Seligman1994,Joyner2012}, whereas those of the NB follow GUE for all symmetry classes. If the spectrum of a QB with $C_3$ symmetry is not separated according to the three subspaces, then its fluctuation properties are described by a composite ensemble, named GOE+2GUE in the following, whose matrices are block diagonal with one GOE block and two GUE blocks of same dimension. For the corresponding NB the composite ensemble consists of three GUE blocks and is denoted by 3GUE. In Ref.~\cite{Zhang2021}, we computed the symmetry-projected eigenstates of massive NBs as described above. For too small masses the eigenvalues corresponding to doublet partners are not degenerate, implying that we do only find agreement of the spectral properties of the NB with those of the DB, GB and HKB around the DPs for sufficiently large mass~\cite{Dietz2020}, even though these exhibit a selective excitation of the two sets of valley states~\cite{Lu2014,Lu2016,Lu2017,Ye2017,Xia2017}. 

\section{The Dirac billiard\label{DB}} 
\begin{figure}[!th]
\includegraphics[width=\linewidth]{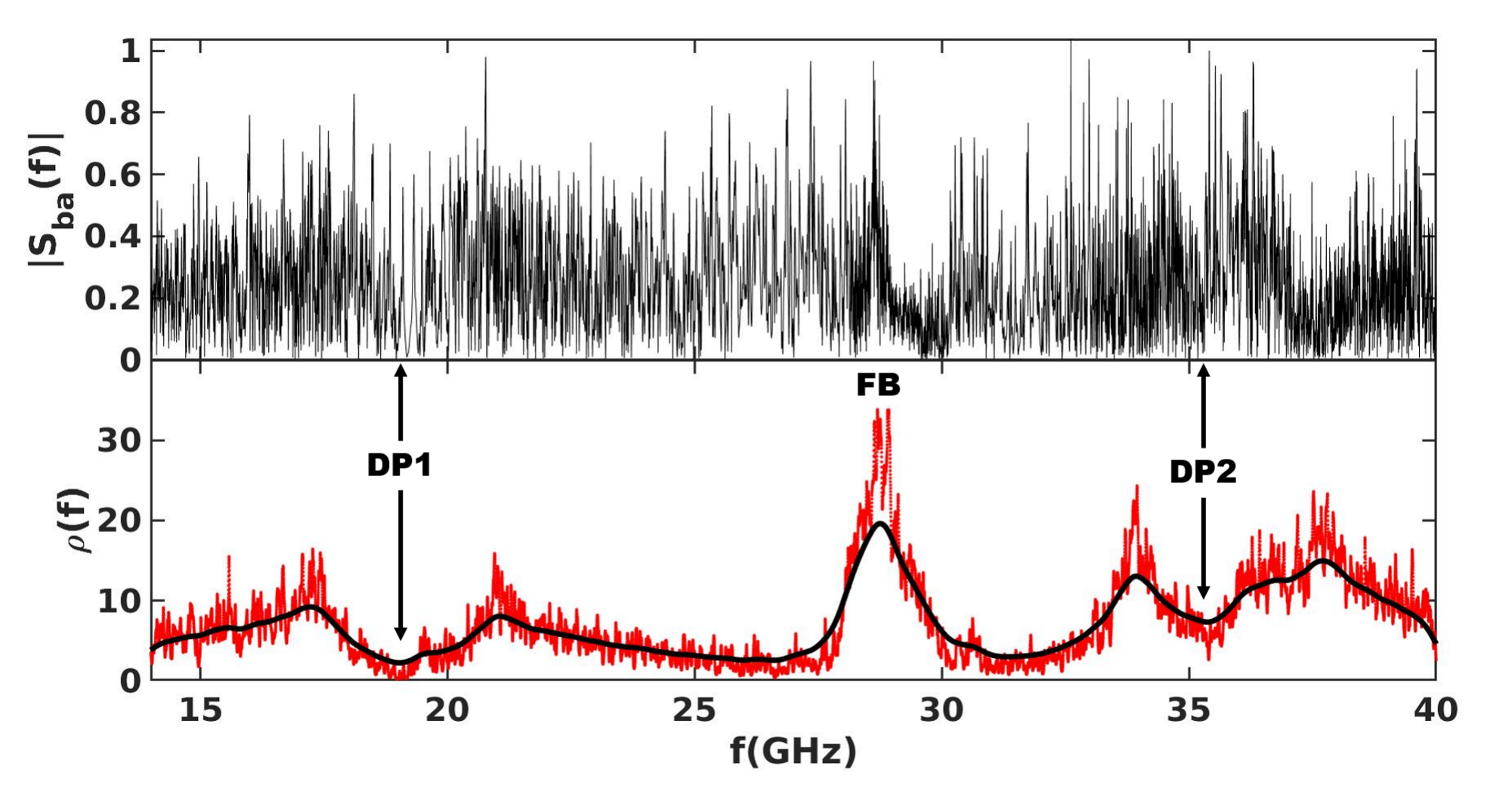} 
	\caption{Upper part: A measured transmission spectrum. The lowest band of propagating modes starts at 13.89 GHz. Lower part: DOS (red) and smoothed DOS (black). The positions of the lower and upper Dirac point (DP1 and DP2) and the FB are indicated.}   
\label{fig:spara}  
\end{figure}

We performed experiments at superconducting conditions. The construction of the DB is explained in the caption of~\reffig{fig:Sketch_Diracbilliard}. The basic ideas are the same as in~\cite{Dietz2015,Dietz2016}.  The cavity consists of a top plate and a basin of 3~mm depth corresponding to a cutoff frequency $f^{cut}=50$~GHz, which contains 1033 metallic cylinders. We chose $r_0=30a_L/\sqrt{3}\simeq 208$~mm in~\refeq{coordinates} with $a_L=12$~mm denoting the lattice constant. The cylinder radius equals $a_L/6$. The sidewall passes through voids, implying Dirichlet BCs at these sites for the corresponding GB. The resonance frequencies were obtained from reflection and transmission spectra. For their measurement we used a Keysight N5227A Vector Network Analyzer (VNA), which sends a rf signal into the resonator at antenna $a$ and couples it out at the same or another antenna $b$ and records the relative phases $\phi_{ba}$ and the ratios of the microwave power, $\frac{P_{out,b}}{P_{in,a}}=|S_{ba}(f)|^2$ yielding the complex scattering matrix element $S_{ba}=|S_{ba}|e^{i\phi_{ba}}$~\cite{Dietz2008,Dietz2009,Dietz2010}. Nine groups of antenna ports consisting of three each, that were positioned such that the $C_3$ symmetry is preserved, were distributed over the whole billiard area, to minimize the possibility that a resonance is missing. This happens when the electric field strength is vanishing at the position of an antenna. The antennas penetrated through holes in the lid into the cavity by about 0.2~mm. The upper part of~\reffig{fig:spara} shows a measured transmission spectrum. Propagating modes are observed above the BE at $f\simeq 13.89$~GHz. 

\begin{figure}[!h]
\includegraphics[width=0.8\linewidth]{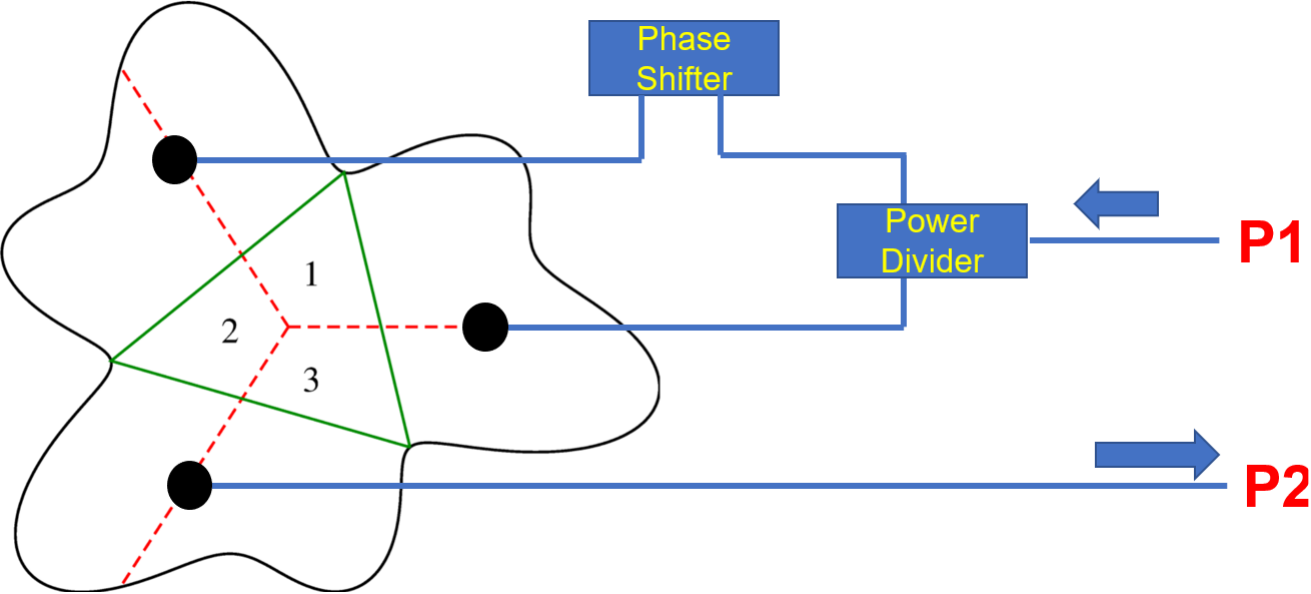}
\caption{Billiards with $C_3$ symmetry can be divided into three fundamental domains that are mapped onto each other under rotation by $\frac{2\pi}{3}$. A possible subdivision is indicated by the red-dashed lines. For the measurements with phase shifters microwaves are fed into the resonator at port P1 of the VNA and split into two signals of equal power and phase by a power divider (GF-T2-20400 with amplitude balance $\lesssim 0.4$~dB and phase balance $\lesssim 5^\circ$), before they are coupled into the resonator via two antennas attached to two ports from one of the nine groups. Their relative phase $\Delta\phi$ is changed by a phase shifter (PE8253 for DC-18.6~GHz and P1507D for 18-26.5~GHz). The microwave power is received through the third antenna port at port P2 of the VNA. This process is irreversible. The shortest connected PO (green lines) has a length of $\tilde{l}_s=11.336r_0/3$.}
\label{fig:sketch}
\end{figure}

The positions of the resonances yield the resonance frequencies. Degeneracies of doublet partners generally are slightly lifted due to experimental imperfection. Consequently finding them can be cumbersome or even impossible, because corresponding resonances overlap. To identify them and to classify them into singlets and doublets we employed a measurement method introduced in~\cite{Dembowski2003} and illustrated in~\reffig{fig:sketch}. When changing the relative phase between the two ingoing signals, the position and shape of the singlets is basically not changed, whereas those of the doublets change considerably, the reason being that they are (nearly) degenerate. Thus a superposition of the associated wave functions (electric-field strength) is excited, whose phases differ~\cite{Zhang2021}. This feature has been used to identify all resonance frequencies in the region of the lower BE and below the DP using the measurements with no phase shifters and with phase shifters. In total 153 measurements were performed for the nine antenna groups, 36 with no power divider and phase shifter, 9 with power divider, and for 6 different relative phases $\Delta\phi$ with two types of phase shifters, namely for frequencies $f\in [13,18]$~GHz with a PE8252 and for $f\in[18,26.5]$~GHz with a P1507D; see~\reftab{tab:2}.
\begin{table}[h!]
        \begin{tabular}{|c|c|c|c|}
                \hline
                Frequency  & Power Div. & Phase Div. & $\#\Delta\Phi\times$ antenna comb. \\ \hline
                13-50GHz   & no           & no           & $1\times 36$              \\ \hline
                13-40GHz   & yes          & no           & $1\times 9$               \\ \hline
                18-26.5GHz & yes          & yes(PE8252)  & $6\times 9$               \\ \hline
                13-18.6GHz & yes          & yes(P1507D)  & $6\times 9$               \\ \hline
        \end{tabular}
        \caption{Measurements were performed for 4 different setups, for different frequency regions with or without power divider and phase shifter and different antenna combinations, as detailed in the table.}
        \label{tab:2}
\end{table}
\begin{figure}[!h]
\includegraphics[width=\linewidth]{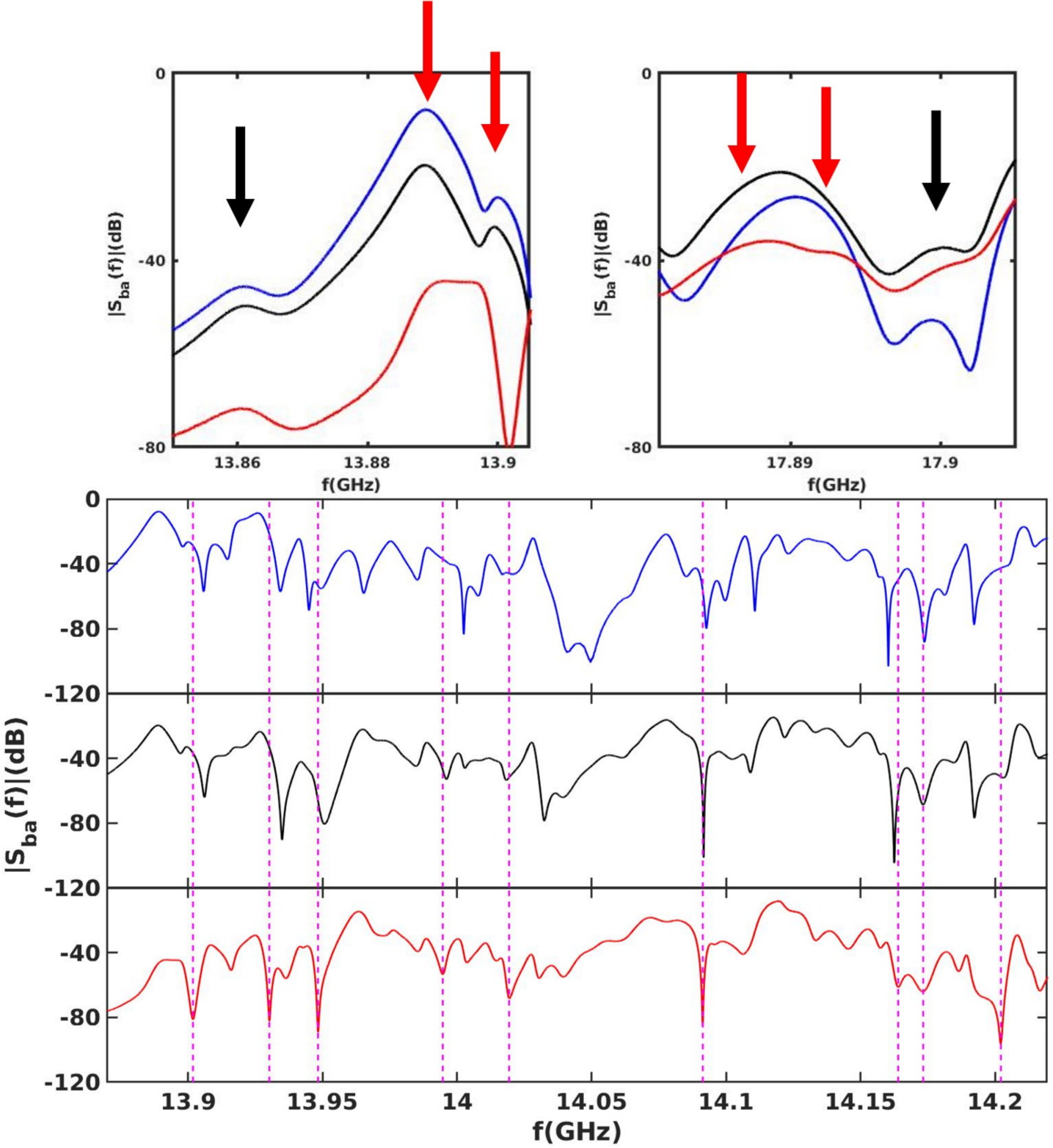}
\caption{Transmission spectra measured with the setup shown in~\reffig{fig:sketch} for relative phases $\Delta\phi=0^\circ$ (red), $\Delta\phi=120^\circ$ (black) and $\Delta\phi=240^\circ$ (blue). The vertical dashed lines are plotted as guidelines to improve the visibility of the changes of the spectra with $\Delta\phi$. The insets to the left and right display the $\Delta\phi$ dependence in zooms into frequency regions comprising one singlet (black arrow) and doublet partners (red arrows).}
\label{fig:spara_phase_shifter}
\end{figure}
Thereby, we were able to identify all resonance frequencies in the region of the lower BE and below the DP1. Even though the quality factor of the resonator was $Q>10^4$ we could not find all resonance frequencies in other regions. 

\begin{figure}[!h]
        \includegraphics[width=0.7\linewidth]{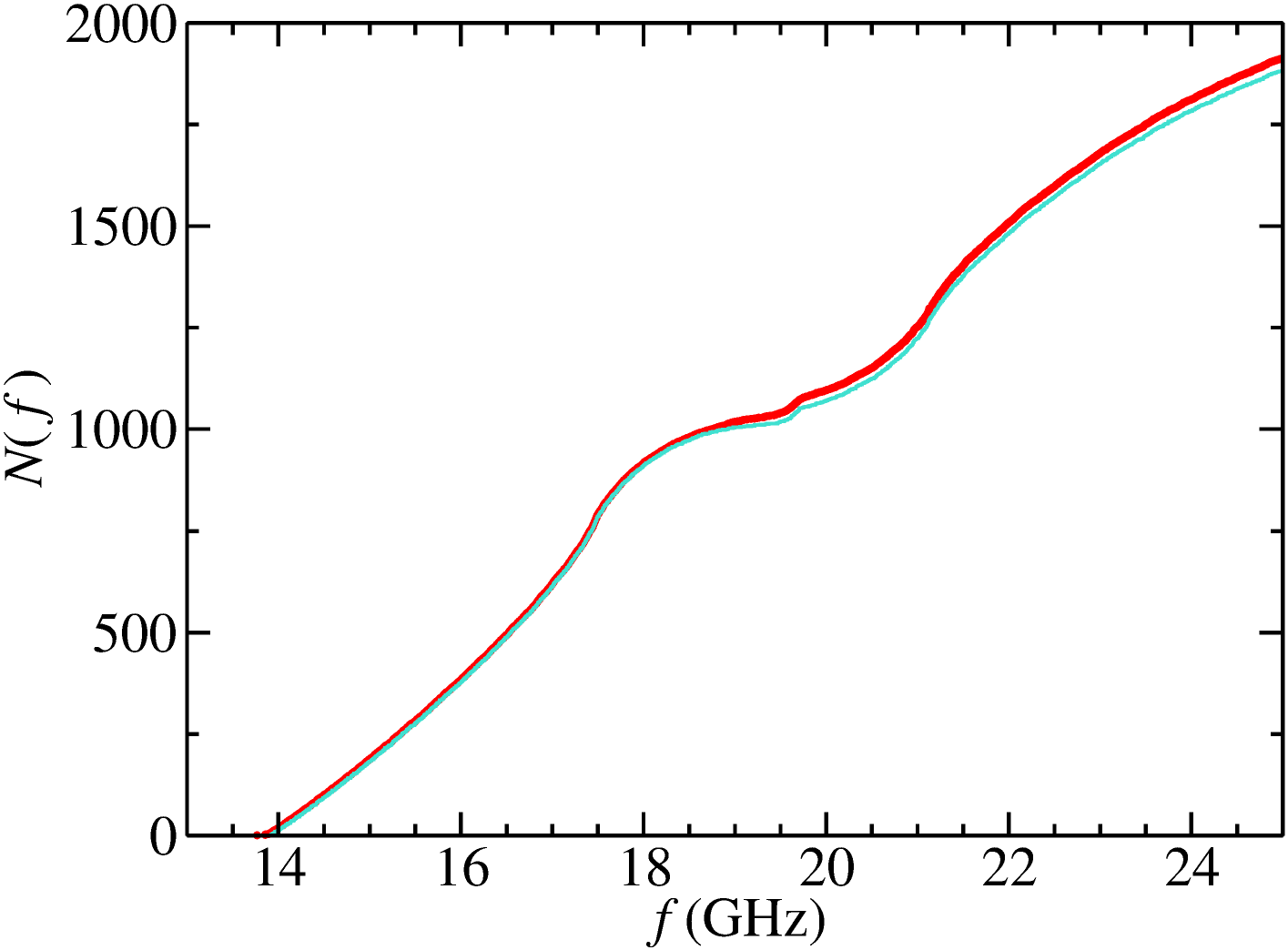}
        \includegraphics[width=0.7\linewidth]{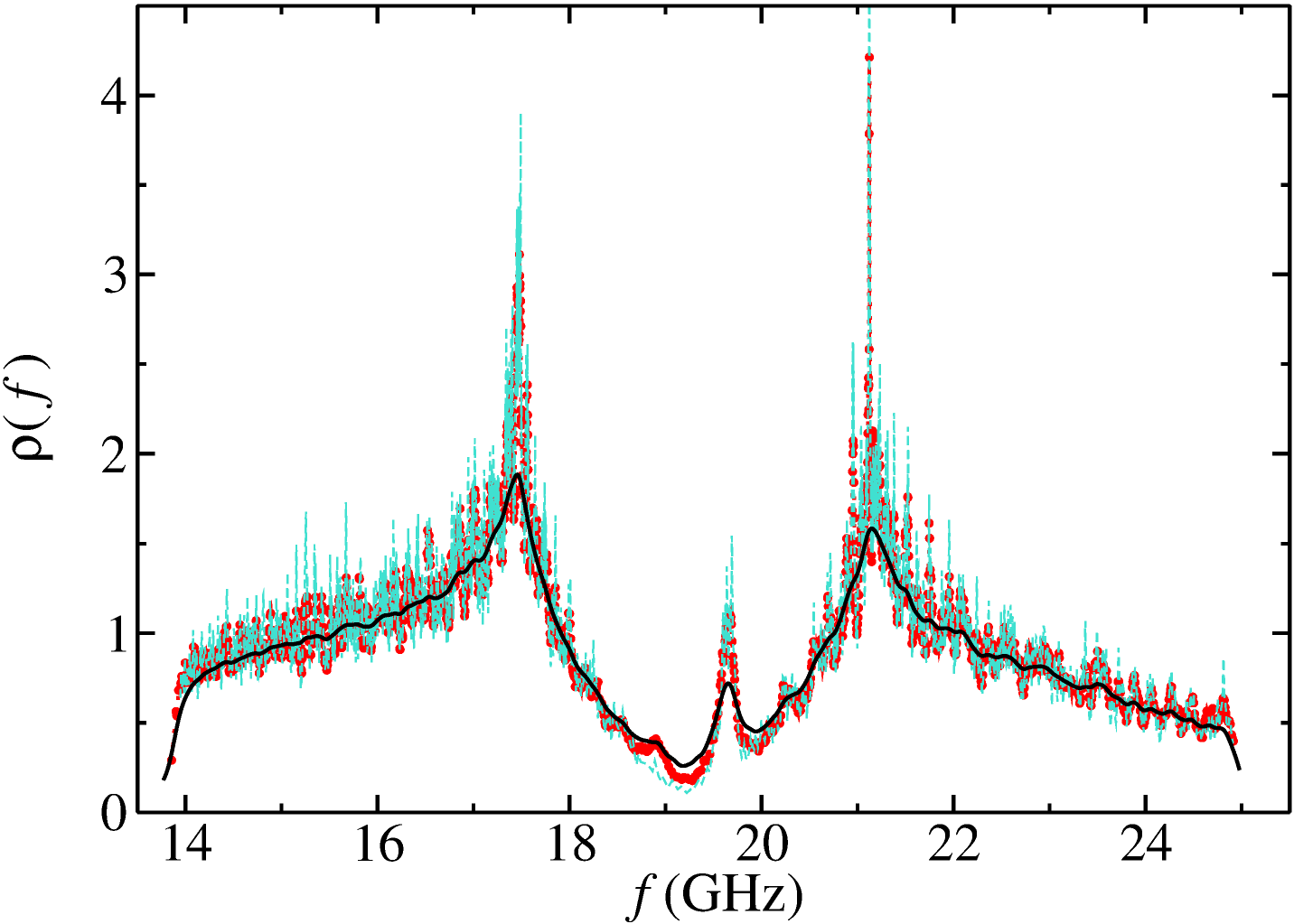}
        \caption{Top: Integrated spectral density obtained from the experimental (red) and with COMSOL computed (turquoise) eigenfrequencies. Bottom: Same as left part for the DOS. The black line shows the smoothed experimental DOS.}
        \label{fig:DOSN}
\end{figure}

In the lower part of~\reffig{fig:spara} we show the density of states (DOS) $\rho(f)$ and the smoothed DOS (black curve). We observe two DPs, denoted by DP1 and DP2, van Hove singularities (VHSs) framing them and a FB. Their frequency values are listed in~\reftab{tab:1} . Around the DP2 the DOS is distorted by an adjacent band~\cite{Zhang2021}. At the FB the resonance frequencies are macroscopically degenerate in a perfect honeycomb-kagome lattice, whereas in the DB degeneracies are slightly lifted due to experimental imperfection and the spreading of the wave-function components located on the sites of the lattice. We, indeed, had to include in the TBM for the HKB couplings and wave-function overlaps~\cite{Reich2002} for up to third-nearest neighbors in the GB sublattice to get agreement with the numerical and experimental DOS~\cite{Dietz2015,Maimaiti2020,Zhang2021}. In the upper part of~\reffig{fig:DOSN}, we compare the integrated spectral densities $N(f)$  obtained from the experimental and computed resonance frequencies. In total 1912 resonance frequencies were identified in that frequency range. The curves start to differ above the lower VHS, which indicates that there not all resonance frequencies were obtained. Note, that at the VHSs the resonance frequencies are nearly degenerate~\cite{Dietz2013}. Similarly, the spectral densities $\rho(f)$, shown in the lower part of~\reffig{fig:DOSN}, agree well except at the VHSs.
\begin{table}[]
\begin{tabular}{|c|c|c|c|}
        \hline
        $f^-_{VHS1}$ & $f_{DP1}$     & $f^+_{VHS1}$ & BG/FB                             \\ \hline
        17.20GHz                      & 19.05GHz       & 21.12GHz                      & $\sim$28.72GHz \\
        \hline
        $f^-_{VHS2}$ & $f_{DP2}$     & $f^+_{VHS2}$ & BG             \\ \hline
        $\sim$33.84GHz                & $\sim$35.42GHz & $\sim$37.52GHz                & $\sim$42.78GHz \\
        \hline
        \end{tabular}
	\caption{Frequencies of the lower ($-$) and upper ($+$) van Hove singularities (VHSs), around the Dirac points (DPs) ($1$) and ($2$), the centers of the band gaps (BGs) and the flat band  (FB) observed in~\reffig{fig:spara}.}
        \label{tab:1}
\end{table}
The  frequency values of the two DPs, denoted by DP1 and DP2, the van Hove singularities (VHSs) framing them and the FB are listed in~\reftab{tab:1}.
\section{Spectral Fluctuations\label{Spectr}}
\begin{figure}[htbp]
\includegraphics[width=0.9\linewidth]{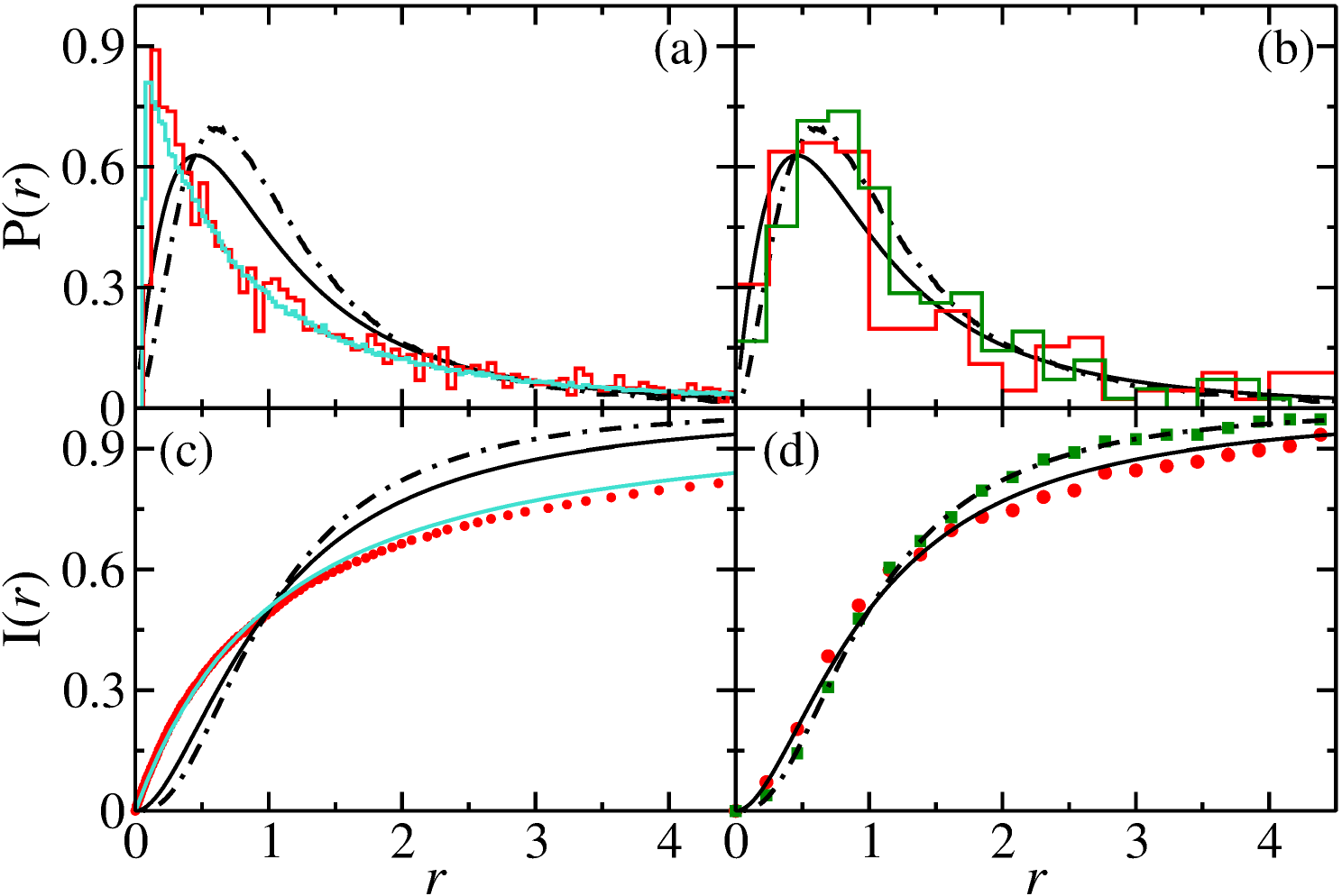}
	\caption{Ratio distributions (upper panel) and cumulative ratio distributions (lower panels). (a), (c): All eigenfrequencies (red histogram and dots) below the FB. (b), (d): Singlets (green histogram and squares) and doublets (red histogram and dots) around the lower BE. The results are compared to those for GOE (solid black lines), GUE (dashed-dotted black lines) and GOE+2GUE (turquoise).}
\label{fig:Ratio}
\end{figure}

The spectral properties were analyzed below the FB in three frequency ranges, namely around the BEs, the VHSs, and in the Dirac region~\cite{Dietz2015,Dietz2016}. These regions are clearly distinguishable in the DOS shown in~\reffig{fig:DOSN}.
\begin{figure}[htbp]
        \includegraphics[width=\linewidth]{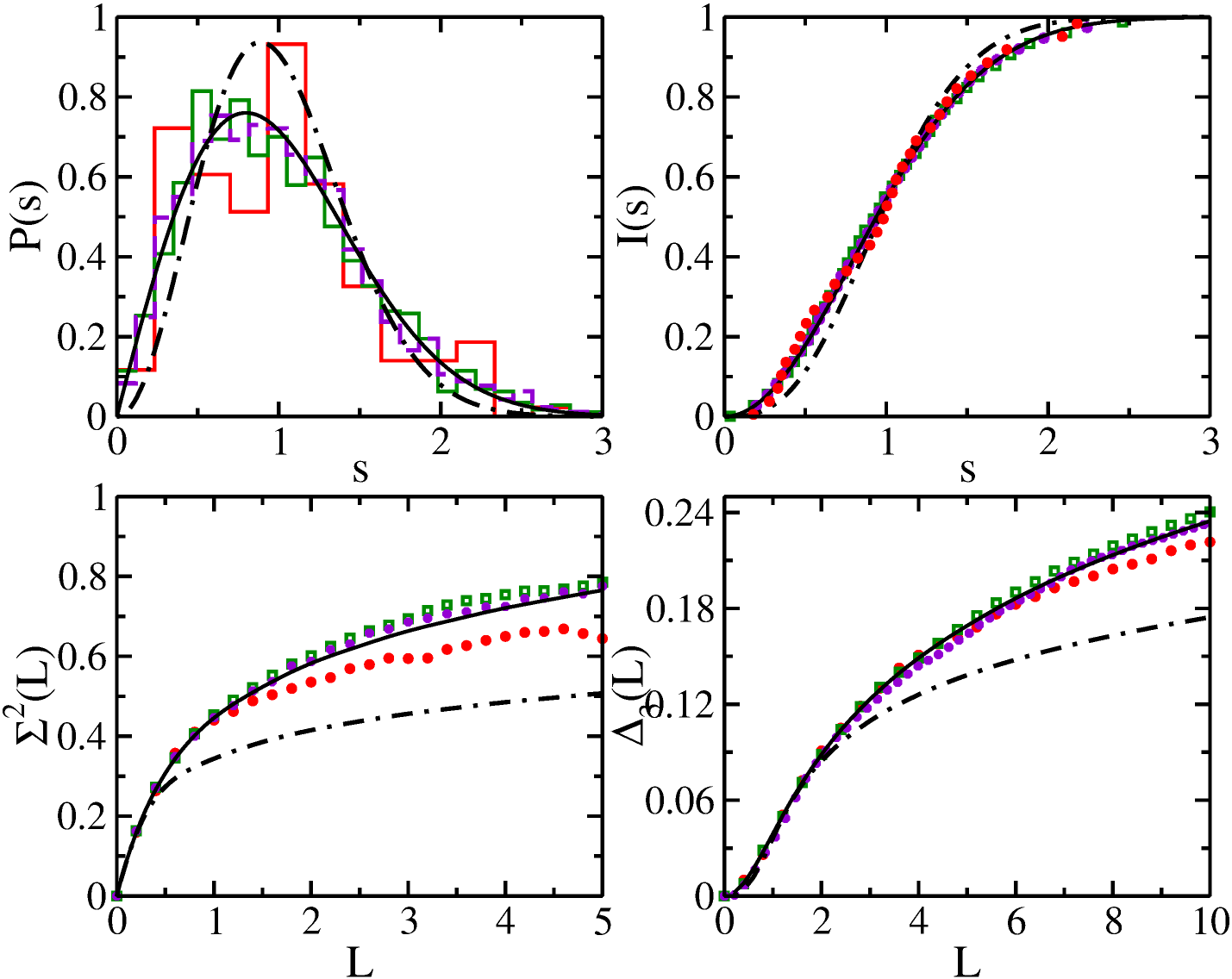}
        \includegraphics[width=\linewidth]{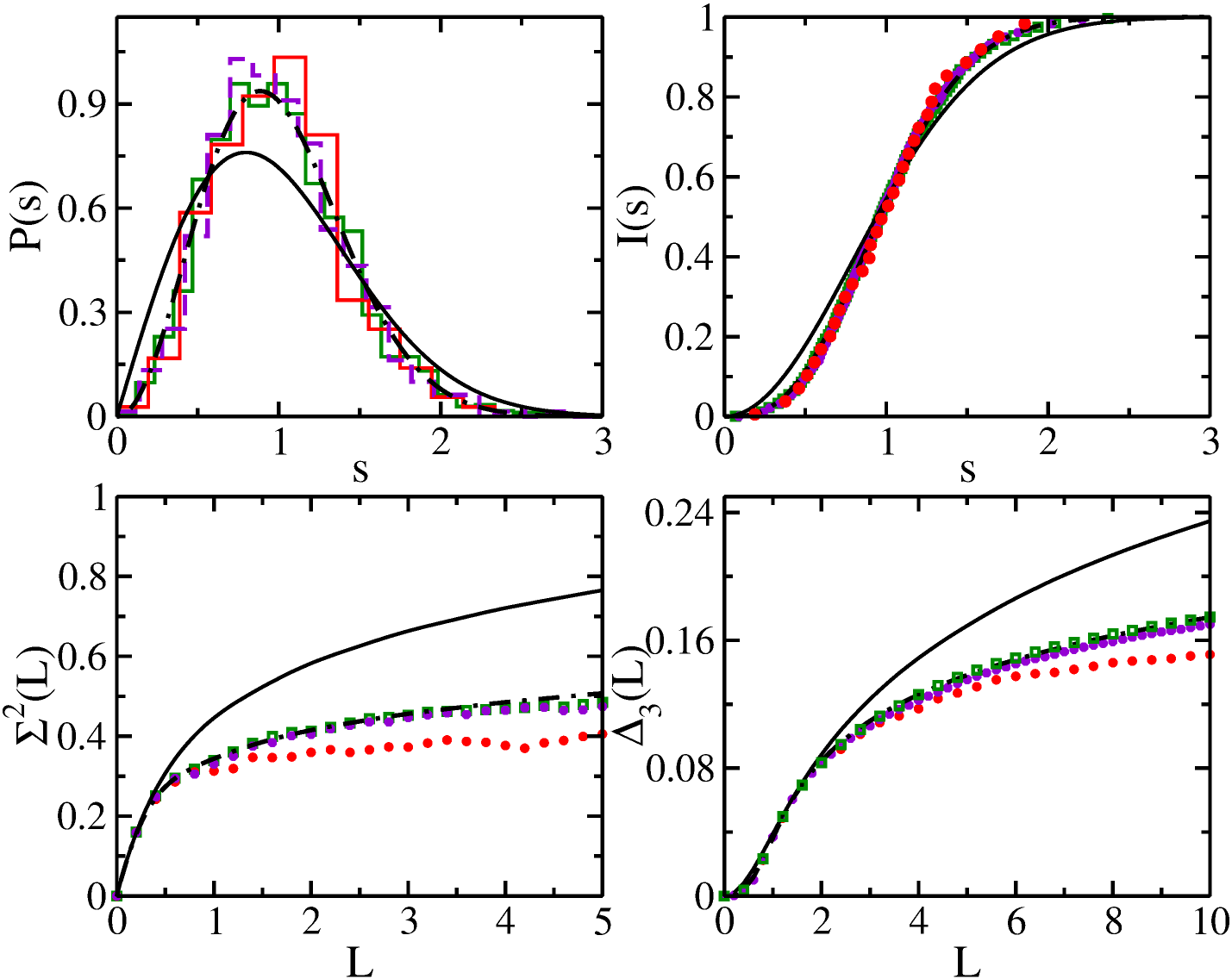}
	\caption{Nearest-neighbor spacing distribution $P(s)$, cumulative nearest-neighbor spacing distribution $I(s)$, number variance $\Sigma^2(L)$ and Dyson-Mehta statistics $\Delta_3(L)$ for the singlets (top) and doublets (bottom) at the lower BE for the DB (red histograms and dots) and GB (green histograms and squares), and the QB (violet dashed-line histograms and stars). The solid and dashed-dot black lines show the curves for GOE and GUE statistics, respectively.}
        \label{fig:spssch}
\end{figure}
\begin{figure}[htbp]
        \includegraphics[width=\linewidth]{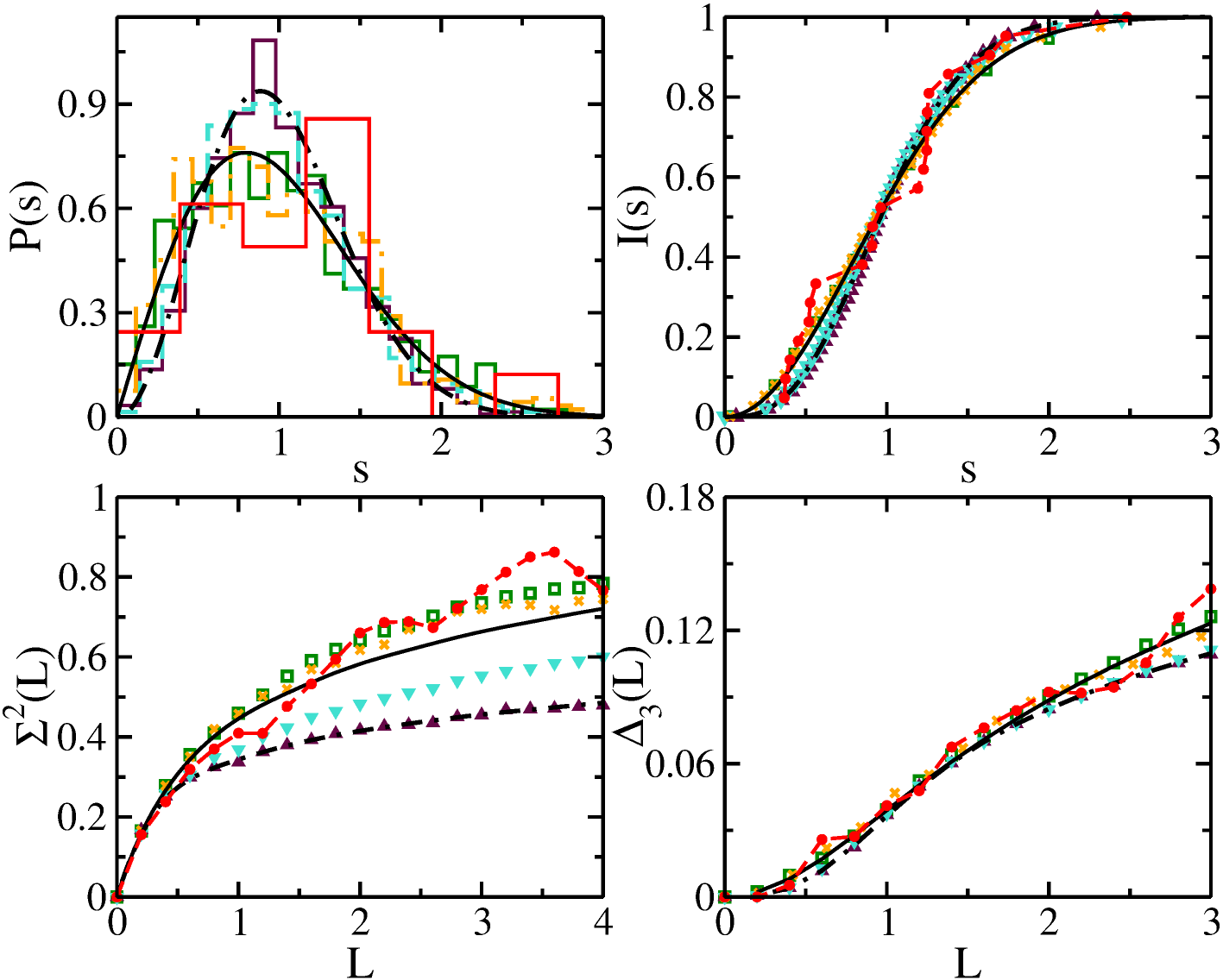}
        \includegraphics[width=\linewidth]{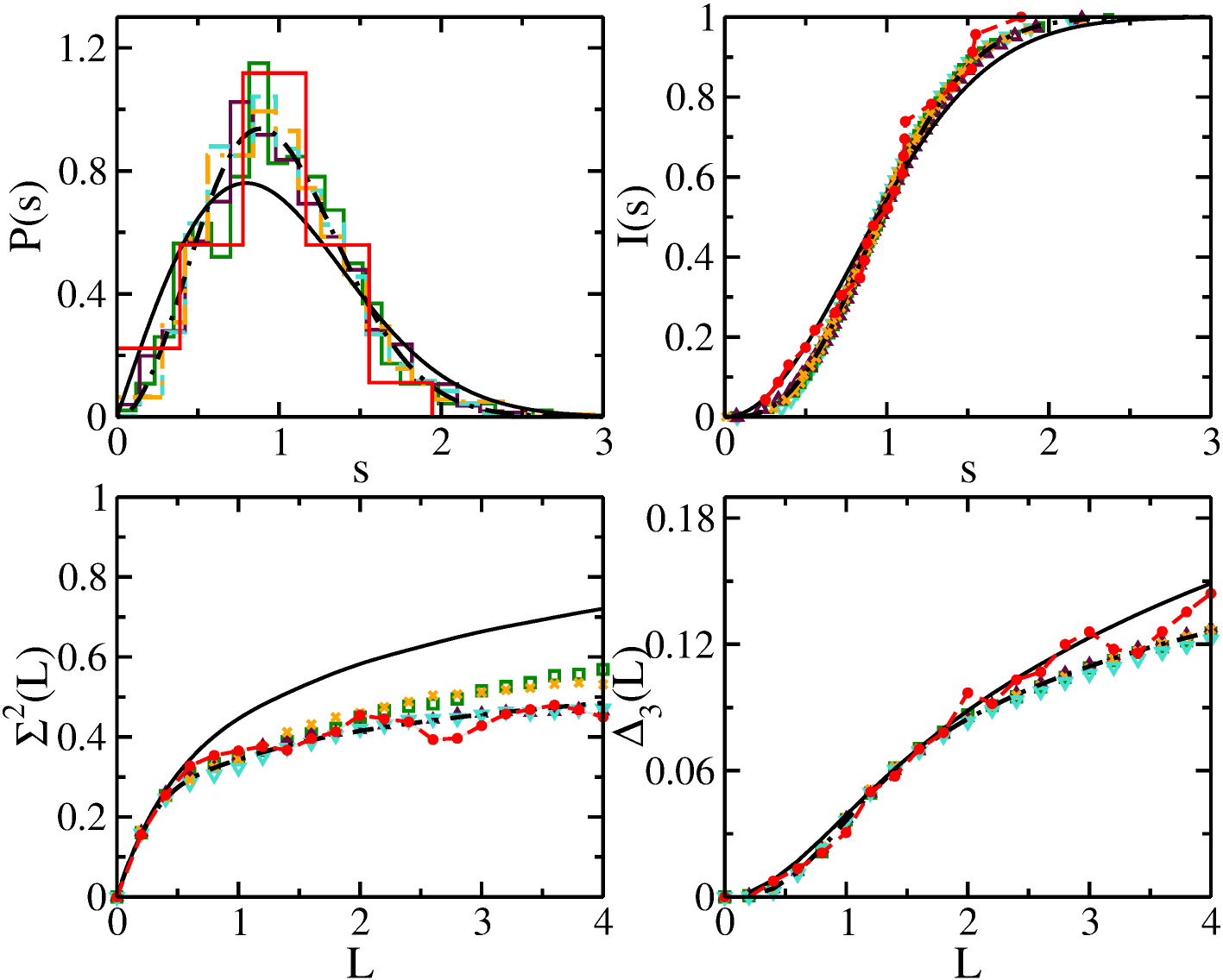}
	\caption{Nearest-neighbor spacing distribution $P(s)$, cumulative nearest-neighbor spacing distribution $I(s)$, number variance $\Sigma^2(L)$ and Dyson-Mehta statistics $\Delta_3(L)$ for the singlets (top) and doublets (bottom) at the DP for the DB (red histograms and dots) and GB (green histograms and squares), and the NB for mass $m_0=0$ (maroon histogram and triangles up), $m_0=20$ (turquoise histograms and triangles down) and $m_0=100$ (orange histograms and crosses). The solid and dashed-dotted black lines show the curves for GOE and GUE statistics, respectively.}
        \label{fig:spsdir}
\end{figure}
We considered 189 levels for each symmetry class starting from the lower BE. Due to the presence of edge states, that lead to the peak observed in the DOS above the DP in~\reffig{fig:DOSN} and yield nonuniversal contributions to the spectral properties~\cite{Dietz2015}, we only considered levels below the DP1, where each subspectrum comprises 26 levels. To unfold the resonance frequencies $f_i$ to average spacing unity, we ordered them by size and determined the number of eigenfrequencies $N(f)$ below $f$. Then we replaced $f_i$ by the smooth part of $N(f)$, $\epsilon_i=N^{smooth}(f_i)$, which we determined by fitting a second order polynomial to $N(f_i)$~\cite{Dietz2015,Zhang2021}. We analyzed the spectral properties in terms of the nearest-neighbor spacing distribution $P(s)$, the integrated nearest-neighbor spacing distribution $I(s)$, the number variance $\Sigma^2(L)$ of $N(f+L)-N(f)$ in an interval of length $L$, and the rigidity of a spectrum of length $L$ $\Delta_3(L)$~\cite{Bohigas1975,Mehta2004}. In ~\reffig{fig:spssch} we show spectral properties of the singlets (top) and doublets (bottom) at the lower BE for the DB (red histograms and dots) and GB (green histograms and squares), and for the QB (violet histograms and stars). They follow the GOE curves (black solid lines) for the singlets and the GUE curves (dashed-dotted black lines) for the doublets in all cases. In~\reffig{fig:spsdir} are plotted the spectral properties of the singlets (top) and doublets (bottom) at the DP1 for the DB (red histograms and dots) and GB (green histograms and squares), and for the NB for mass $m_0=0$ (maroon histograms and triangles up), $m_0=20$ (turquoise  histograms and triangles down) and $m_0=100$ (orange histograms and crosses). For the DB and the GB we find the same behavior as around the lower BE, whereas for the NB with $m_0=0$ the spectral properties agree with GUE for the singlets and doublets, and are between GUE and GOE for the singlets for $m_0=20$. For $m_0=100$ the spectral properties agree well with those of the corresponding QB, that is, there the nonrelativistic limit is reached. Deviations may be attributed to the small number of levels and to the presence of short periodic orbits~\cite{Zhang2021}. The shortest connected one is shown in~\reffig{fig:sketch}. We, in addition, considered the distribution $P(r)$ and the cumulative distribution $I(r)$ of the ratios~\cite{Oganesyan2007,Atas2013} $r_i=\frac{\epsilon_{i+1}-\epsilon_i}{\epsilon_{i}-\epsilon_{i-1}}$, which are dimensionless so that unfolding is not needed~\cite{Dietz2016,Maimaiti2020}. The results for all resonance frequencies below the FB are shown in the left part of~\reffig{fig:Ratio}, those of the singlets (red) and doublets (green) at the lower BE in the right part. The former are compared to those of random matrices from the GOE+2GUE. In all, the spectral properties agree well with those obtained from the COMSOL Multiphysics computations in~\cite{Zhang2021} and with random-matrix theory (RMT) predictions for nonrelativistic QBs with $C_3$ symmetry.

\section{$S$-matrix Fluctuations\label{Scatt}} 
\begin{figure}[htbp]
        \includegraphics[width=\linewidth]{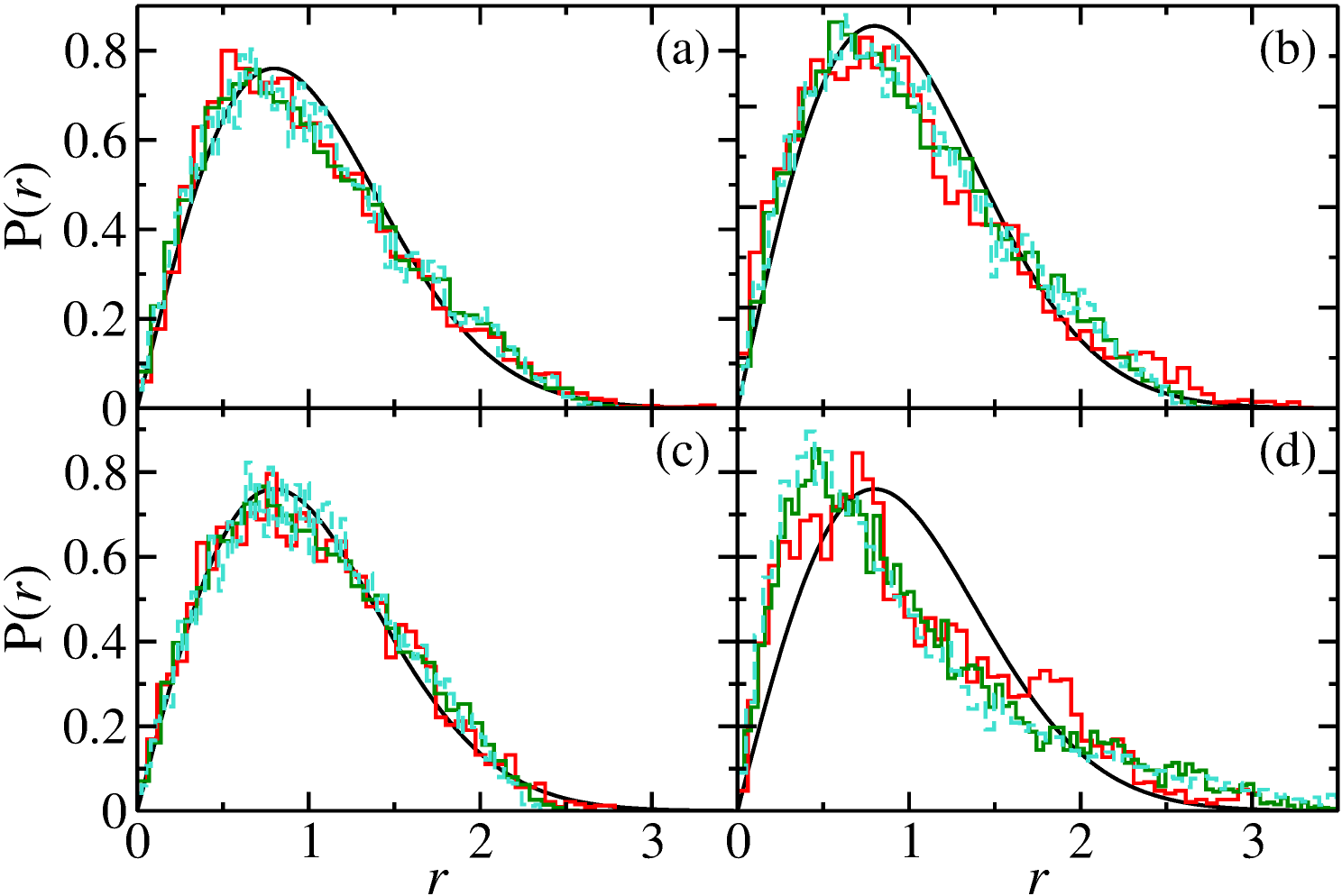} 
	\caption{Distributions of the transmission amplitudes $r=\vert S_{12}\vert/\langle\vert S_{12}\vert\rangle$ (red histogram) in (a) the region around the lower band edge $f\in[15,16]$~GHz for antennas 4, 5 and 6, (b) same as (a) for the upper band edge $f\in[23,24]$~GHz, (c) around the lower VHS $f\in[17.3,17.6]$~GHz for antennas 1 and 2 and (d) the same as (c), but for the upper VHS $f\in[21,21.3]$~GHz. They are compared to distributions obtained from the RMT model~\refeq{eqn:Sab} with $\hat H$ from the GOE+2GUE (green histograms) and 3GUE (turquoise histograms). Best fit is found for $T_1=0.57, T_2=0.55$ and (a) $\tau_{abs}=1.0$, (b) $\tau_{abs}=0.8$, and for $T_1=0.67, T_2=0.69$ and $\tau_{abs}=1.0$ (c). In (d) we use corresponding PLBMs with $\alpha =0.3$ and otherwise the same values as in (c). The black solid lines exhibit the bivariate Gaussian expected in the Ericson regime.
}\label{VertBEVH}
\end{figure}

\begin{figure}[htbp]
\includegraphics[width=\linewidth]{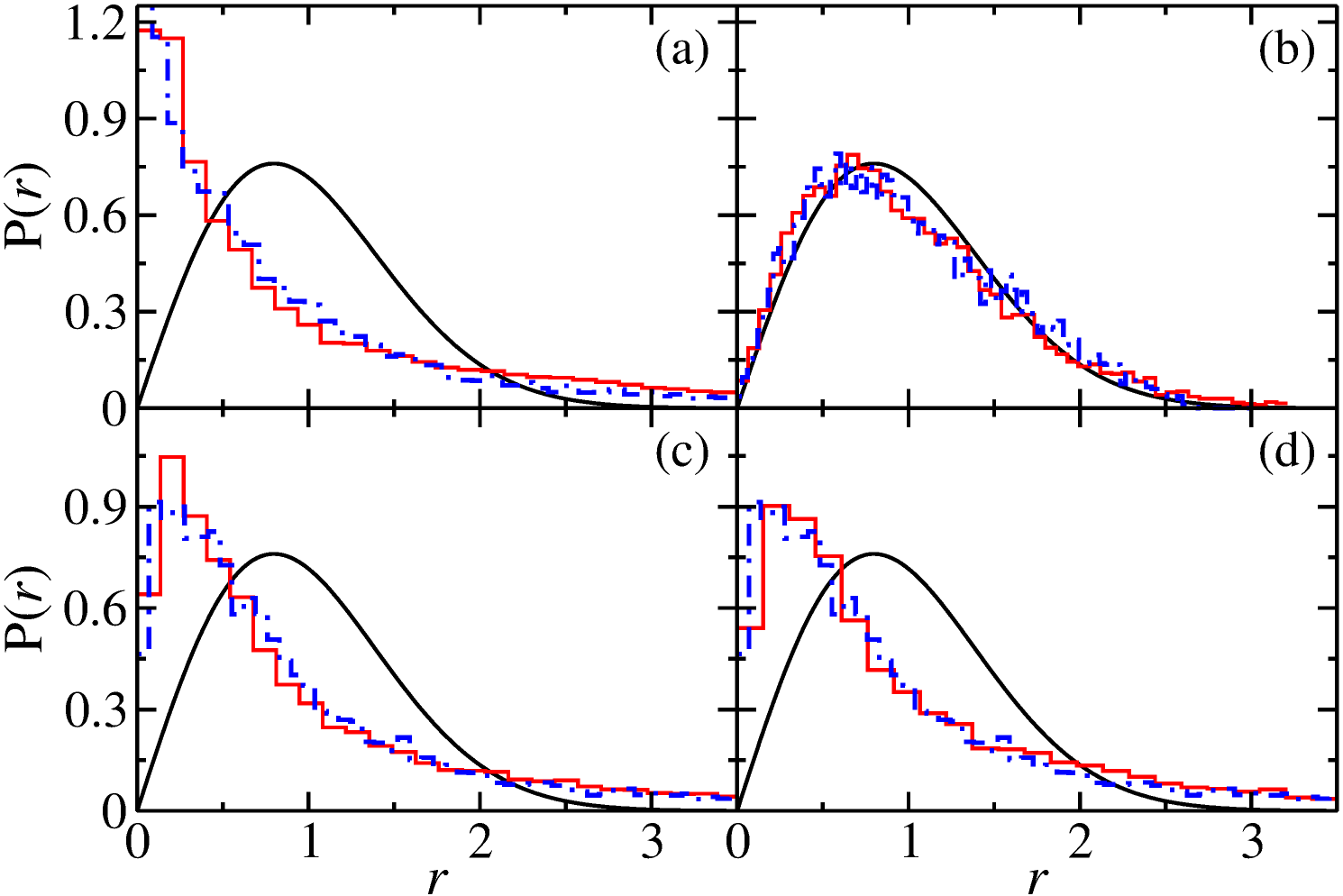} 
\includegraphics[width=\linewidth]{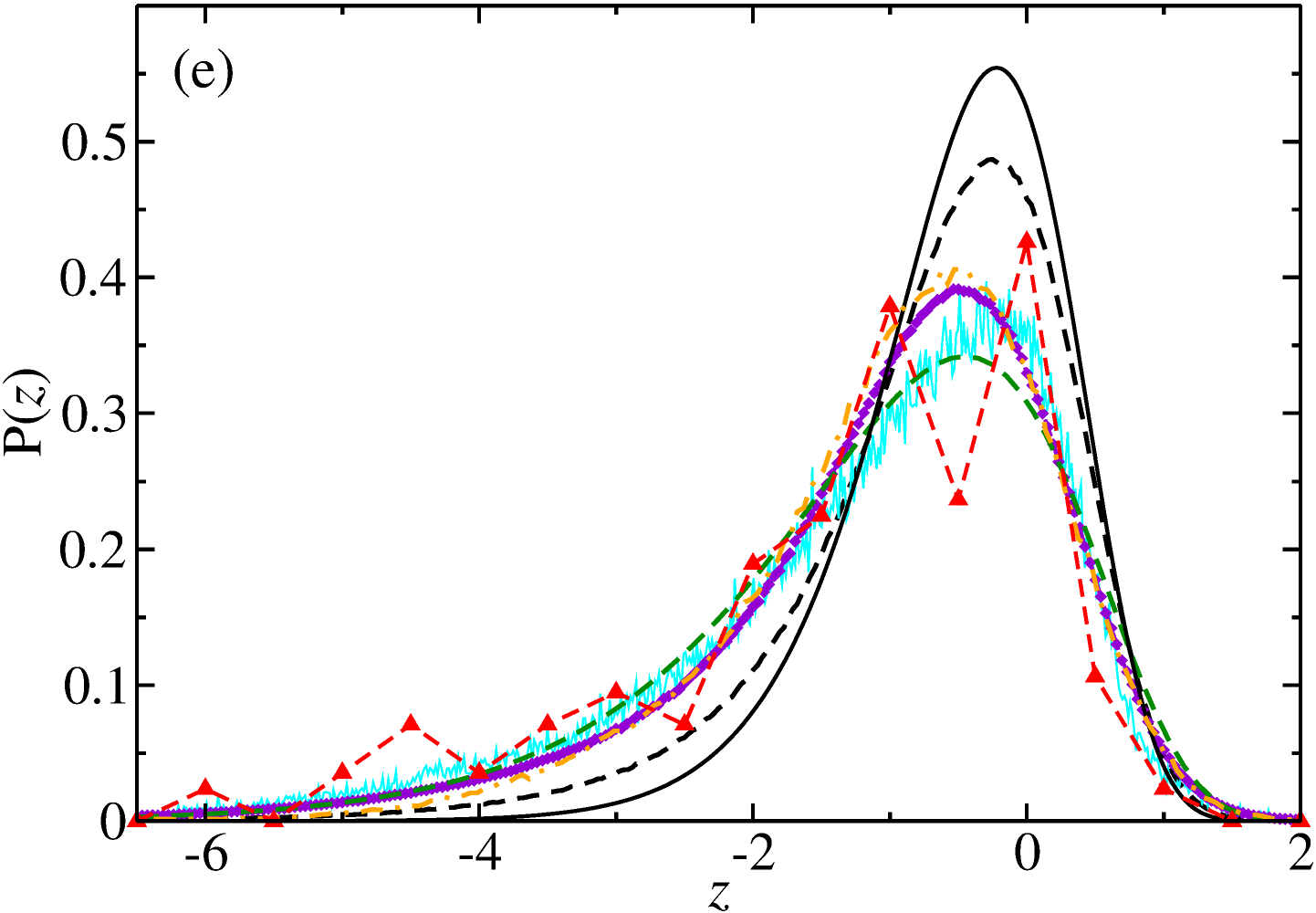}
	\caption{(a)-(c): Distributions of the transmission amplitudes $r=\vert S_{12}\vert/\langle\vert S_{12}\vert\rangle$ (red histograms) measured in the FB $f\in[28,29]$~GHz with all antennas (a),  with antennas 1 and 2 (b), with antennas 3, 4, 5 and 6 (c) and with antennas 7, 8 and 9 (d). They are compared to the RMT model~\refeq{eqn:Sab} with the PLBMs (blue histograms) generated from random matrices from the GOE+2GUE for $T_1=0.67, T_2=0.69, \tau_{abs}=1.0$ and $\alpha =1.0$ (a), $\alpha =0.1$ (b), and $\alpha =0.7$ in (c) and (d). The black solid lines exhibit the bivariate Gaussian expected in the Ericson regime. (e) Strength distribution in the Dirac region $f\in [18.4,19.1]$ (red triangles) obtained from antenna groups 7, 8 and 9, and from the computed wave functions of the GB in the same outer region (cyan line). They are compared to the analytical results for GOE (green dashed line), 3GUE (black solid line), GOE+2GUE (black dashed line) and to RMT simulations for PLBMs with $\alpha\simeq 0.7-0.8$ for 3GUE (orange dashed-dotted line) and GOE+2GUE (violet diamonds).}  
\label{VertVHFB}   
\end{figure}

We also investigated fluctuation properties of the $S$ matrix associated with the measurement process and compared them to RMT predictions for quantum-chaotic scattering systems derived from the scattering matrix approach~\cite{Mahaux1969} which was developed in the context of compound nuclear reactions and extended to microwave resonators in~\cite{Albeverio1996},
\begin{equation}
        S_{ba}(f) = \delta_{ba} - 2\pi i\left[\hat W^\dagger\left(f\II-\hat H^{eff}\right)^{-1}\hat W\right]_{ba}.
\label{eqn:Sab}
\end{equation}
Here, $\hat H^{eff}=\hat H-i\pi\hat W\hat W^\dagger$ with $\hat H$ modeling the universal spectral properties of the DB.  Since we did not separate the resonance spectra by symmetry, we chose for $\hat H$ random matrices from the composite ensemble GOE+2GUE and from the 3GUE for comparison. The matrix elements of $\hat W$ are real, Gaussian distributed with $W_{a \mu}$ and $W_{b \mu}$ describing the coupling of the antenna channels to the resonator modes. Furthermore, we chose $\Lambda$ equal fictituous channels to account for the Ohmic losses in the walls of the resonator~\cite{Dietz2009,Dietz2010}. Direct transmission between the antennas was negligible, so that the frequency-averaged $S$-matrix was diagonal, implying that $\sum_{\mu = 1}^N W_{e \mu} W_{e^\prime \mu}=N v_{e}^2 \delta_{ee^\prime}$~\cite{Verbaarschot1985}. The parameters $v^2_{e}$ denote the average strength of the coupling of the resonances to channels $e$. For $e=a,\ b$ they correspond to the average size of the electric field at the position of the antennas $a$ and $b$ and they yield the transmission coefficients $T_{e} = 1 - \vert\left\langle{S_{ee}}\right\rangle\vert^2$, which are experimentally accessible~\cite{Dietz2010}. Actually, $v_e$ and $\tau_{abs}=\Lambda T_c$ are the input parameters of the RMT model~\refeq{eqn:Sab} where they are assumed to be frequency independent. This is fulfilled because we analyzed data in windows of size $\leq 1$~GHz~\cite{Dietz2010}. We considered three parts of the DB, defined by the location of the antennas $a$ and $b$, namely an inner region (groups 1, 2) around the center of the billiard domain, a middle region (groups 3, 4, 5, 6) and an outer region (groups 7, 8, 9); see~\reffig{fig:Sketch_Diracbilliard}. In~\reffig{VertBEVH} distributions of the rescaled transmission amplitudes are shown around the lower (a) and upper (b) BE, and around the lower (c) and upper (d) VHS. At the BEs the distributions do not depend on the positions of the antennas and are well described by the RMT model~\refeq{eqn:Sab} both for the GOE+2GUE (green) and the 3GUE (turquoise) case which, actually, are barely distinguishable. There the wave-functions are similar to those of the corresponding QB~\cite{Zhang2021}. For the lower VHS and for the FB, shown in~\reffig{VertVHFB}, we only find good agreement with the RMT results for the inner group. Otherwise we do not find any agreement around the VHSs and FB. Instead, these distributions are well described by the $S$-matrix model~\refeq{eqn:Sab} when using power-law banded random matrices (PLBM)~\cite{Mirlin1996}, obtained by multipling the off-diagonal elements $H_{ij}$ of $\hat H$ by a factor $\vert i-j\vert^{-\alpha}$. This ensemble interpolates between localized ($\alpha \gtrsim 1$) and extended ($\alpha =0$) states. This is demonstrated in~\reffig{VertBEVH} (d) and in~\reffig{VertVHFB} (a)-(d). Thus these deviations may be attributed to localization of the electric-field intensity in parts of the DB.

\begin{figure}[h!]
\includegraphics[width=\linewidth]{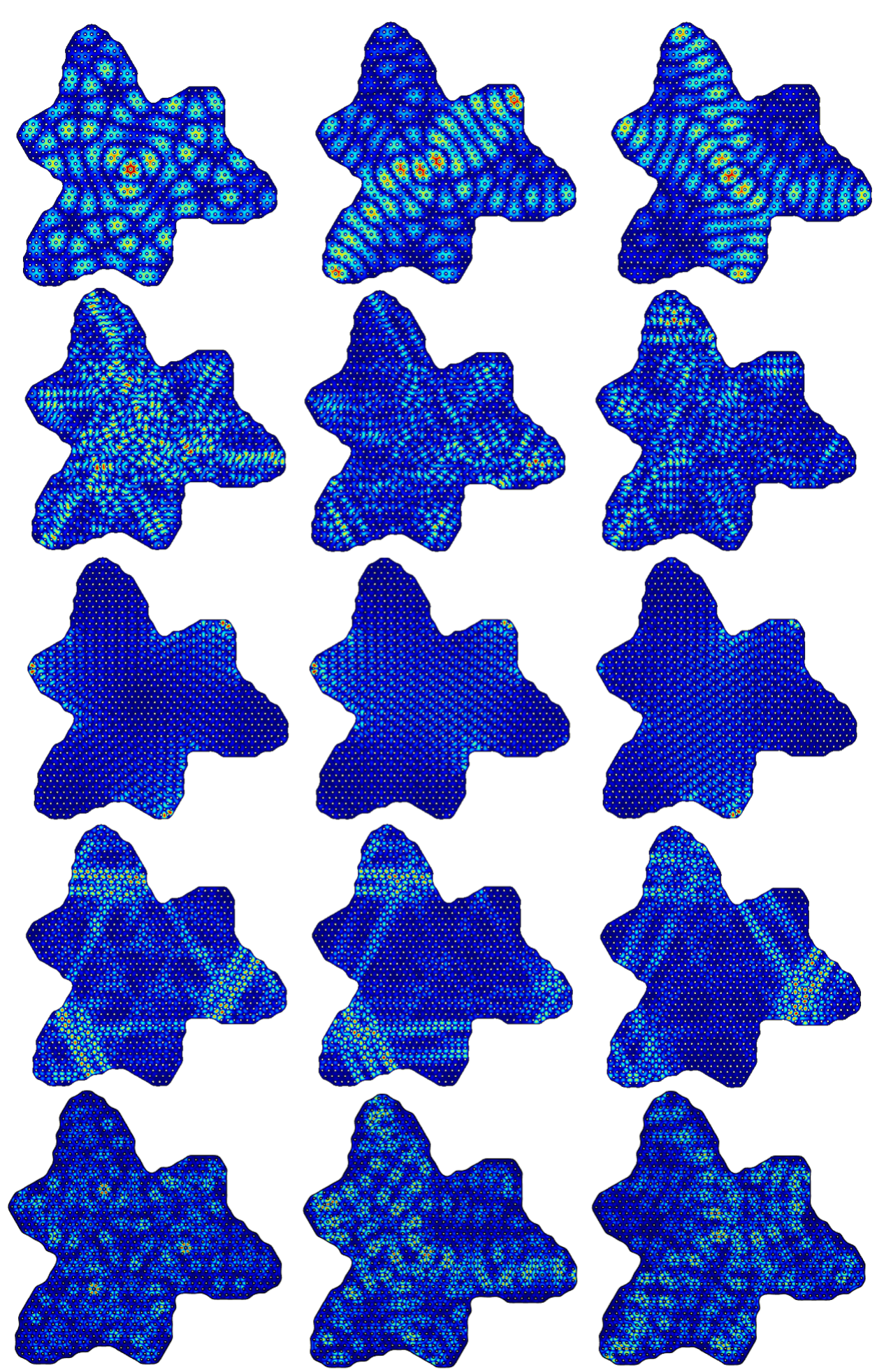}
	\caption{Computed distributions of the electric-field intensity of the DB of the singlets (first column) and doublets (second and third column) corresponding to state number $n$ with resonance frequency $f_n$ in the region below the flat band for, from top to bottom, the lower BE, the lower VHS, the lower DP and the upper VHS. The first row shows, from left to right, examples for $n=94, f_n=14,46$~GHz, $n=104, f_n=14.52$~GHz, $n=105, f_n=14.53$~GHz, the second row for $n=716, f_n=17.224$~GHz, $n=717, f_n=17.232$~GHz, $n=718, f_n=17.235$~GHz, the third row for $n=1007, f_n=19.03$~GHz, $n=1008, f_n=19.073$~GHz, $n=1009, f_n=19.073$~GHz, the fourth row for $n=1252, f_n=21.066$~GHz, $n=1253, f_n=21.072$~GHz, $n=1254, f_n=21.072$~GHz and the fifth row for $n=2560, f_n=28.7503$~GHz, $n=2561, f_n=28.7506$~GHz, $n=2562, f_n=28.7507$~GHz.}
\label{fig:WFDB}
\end{figure}

\begin{figure}[!h]
\includegraphics[width=\linewidth]{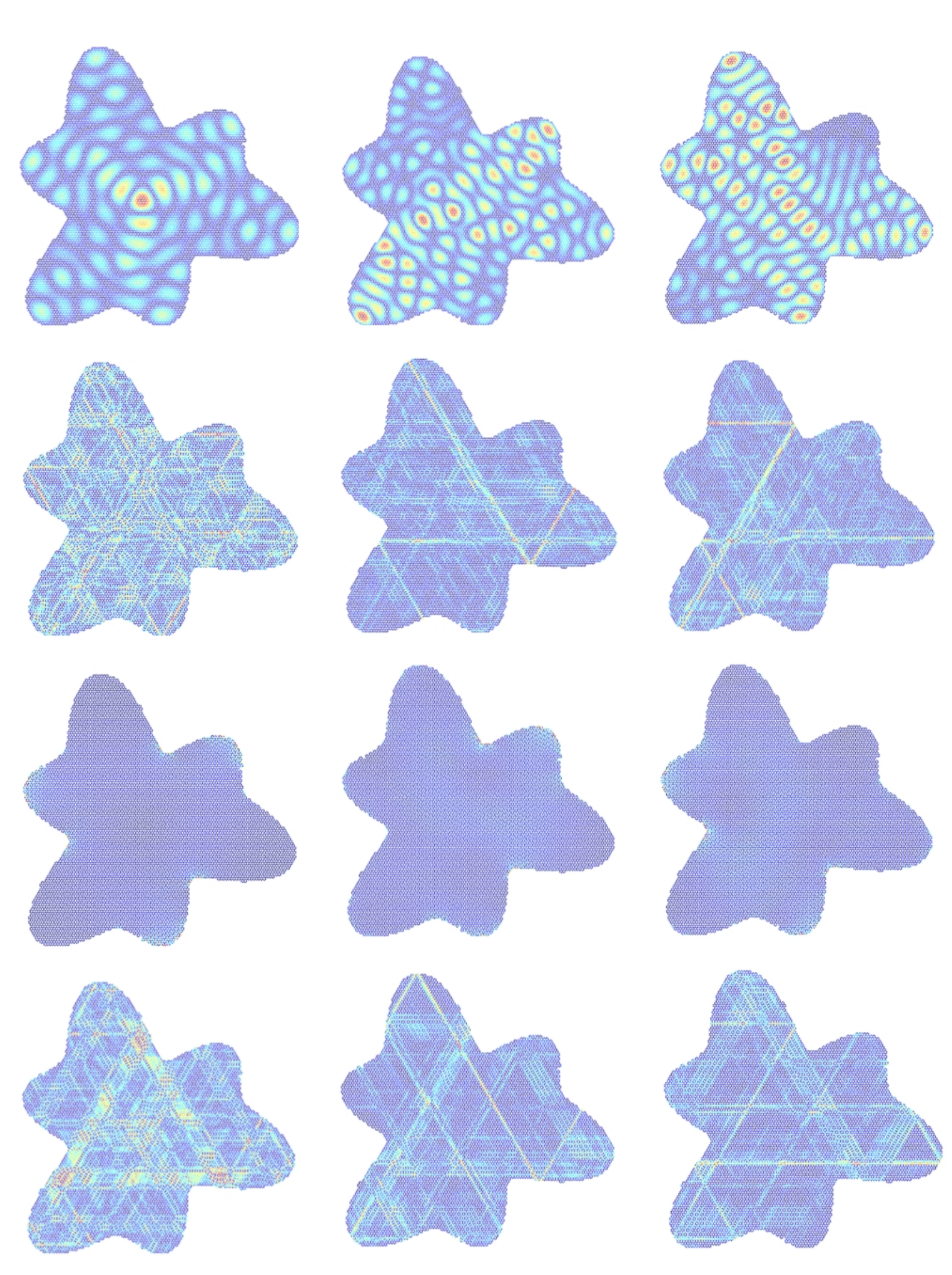}
\caption{Computed distributions of the wave-function intensity of the singlets (first column) and doublets (second and third column) of the GB corresponding to state number $n$ for, from top to bottom, the lower BE, the lower VHS, the lower DP1 and the upper VHS. The first row shows, from left to right, examples for $n=94,164,165$, the second row for $n=5897,5918,5919$, the third row for $n=7981,7982,7983$, the fourth row for $n=10011,10013,10014$.}
\label{fig:WFGB}
\end{figure}

In~\reffig{fig:WFDB} we show typical intensity distributions of the electric field strength of the DB and of the wave functions of the corresponding GB in~\reffig{fig:WFGB}. Examples are shown for the region below the flat band for singlets (first column) and corresponding doublets (second and third comlumn), from top to bottom, around the lower BE (first row), around the lower VHS (second row), around the DP (third row), around the upper VHS (fourth row) and for the DB also in the region of FB (fifth  row). The wave functions of the doublet partners are superpositions of the corresponding symmetry-projected states with $l=1,2$, and thus their intensity distributions exhibit different patterns. Around the BE, the intensity distributions are mostly spread over the whole billiard domain, some are localized around shortest periodic orbits, e.g., the connected one is shown in~\reffig{fig:sketch}, whereas around the VHS we observe especially for the upper VHS a strong localization around periodic orbits in the bulges of the billiard. Accordingly we observe deviations from RMT predictions in the corresponding $S$-matrix amplitude distributions for the middle and outer antenna groups, whereas good agreement is found when using power-law banded random matrices in Eq. (3). Note, that the amplitudes of the resonances depend on the electric-field strength at the position of the measuring antennas, and their distributions are obtained from averaging over all symmetry classes.

At the DP the resonances are well isolated. Therefore, in that region we can obtain information on the properties of the wave-function components in terms of the strength distribution~\cite{Dembowski2005}. Namely, for sufficiently isolated resonances the $S$-matrix has the form
\begin{equation}
S_{a b} = \delta_{a b}
    - i\frac{\sqrt{\Gamma_{\mu a}\Gamma_{\mu b}}}
             {f-f_{\mu} + \frac{i}{2}\Gamma_{\mu}
            }
\label{SMatrix}
\end{equation}
close to the $\mu$th resonance frequency $f_{\mu}$ with $\Gamma_{\rm\mu}$ denoting the total width of the corresponding resonance~\cite{Alt1995}. The partial widths $\Gamma_{\mu a}$ and $\Gamma_{\mu b}$ are proportional to the electric-field intensities at antennas $a$ and $b$. They cannot be determined individually, however, the strengths $z=\Gamma_{\mu a}\Gamma_{\mu b}$ may be obtained with high precision by fitting this expression to the resonances~\cite{Dembowski2005}. The strength distribution corresponds to the distribution of the products of the squared moduli of two wave-function components in the DB, or of two eigenvector components for the associated RMT model~\cite{Dietz2006a,Guhr1998}. For 3GUE it coincides with that of GUE, $P^{GUE}(z)=2K_0(2\sqrt{z})$, that of GOE+2GUE is given by $P^{GOE+2GUE}(z)=\frac{1}{3}\left[P^{GOE}(z)+2P^{GUE}(z)\right]$, where $P^{GOE}(z)=K_0(\sqrt{z})/(\pi\sqrt{z})$. Here, $K_0(z)$ denotes the modified Bessel function of order zero. In~\reffig{VertVHFB} (e) we compare these analytical expresssions to the distributions obtained for the DB in the Dirac region (red triangles). However, like for the FB we find only agreement with the RMT distributions, when using the corresponding PLBM $\alpha\simeq 0.7-0.8$, where that for GOE+2GUE (violet diamonds) is better than that for 3GUE (orange dashed-dotted lines).  

\section{Conclusions\label{Concl}}
We performed experiments with a superconducting DB, whose shape has a $C_3$ symmetry. To identify the resonance frequencies and to separate them into the three symmetry classes we successfully employed a procedure, which was originally developed for hollow microwave billiards~\cite{Dembowski2003}, thereby demonstrating that it is applicable even to complex structures like the DB. We confirm results which were obtained in Ref.~\cite{Zhang2021} from numerical computations, namely, find good agreement of the spectral properties with those of the QB, GB and HKB of corresponding shape, and with those of massive relativistic NBs only beyond a certain mass. We also investigated properties of the wave functions below the DP1, where the DOS is low, in terms of the strength distribution. We find good agreement with the corresponding distributions of random matrices from GOE+2GUE when using PLBMs, corroborating that the wave functions are localized~\cite{Bittner2012}. Yet, the spectral properties of the associated resonance frequencies agree well with those of typical quantum systems with a $C_3$ symmetry and a chaotic classical dynamics. Furthermore, we for the first time investigated the fluctuation properties of the measured $S$ matrix in the regions around the BEs, the VHSs and the FB, which are not accessible numerically. In the nonrelativistic regime we find good agreement with those of the RMT model~\refeq{eqn:Sab} for GOE+2GUE, whereas for the other regions we took account of the localization observed in parts of the DB by using PLBMs. Around the VHSs the ratio distributions agree well with those of random matrices from the GOE+2GUE. From these observations we may conclude that even in regions, where the wave functions are localized in parts of the DB, the spectral properties comply with those of typical quantum systems whose corresponding classical dynamics is chaotic~\cite{Dietz2016}.

\section{Acknowledgement} This work was supported by the NSF of China under Grant Nos. 11775100, 12047501 and 11961131009. WZ acknowledges financial support from the China Scholarship Council (No. CSC-202106180044). BD and WZ acknowledge financial support from the Institute for Basic Science in Korea through the project IBS-R024-D1.

\bibliography{References}
\end{document}